\documentclass[aps,prb,twocolumn,floatfix,showpacs,superscriptaddress]{revtex4-1}
\usepackage{epsfig}

\begin{document}

\newcommand{\bra}[1]{\left\langle #1 \right|}
\newcommand{\ket}[1]{\left| #1 \right\rangle}
\newcommand{\sd}{\downarrow}
\newcommand{\su}{\uparrow}

\title{Analysis of the Kondo effect in ferromagnetic atomic-sized contacts}

\author{M.~R. Calvo}
\affiliation{
  Departamento de F{\'i}sica Aplicada, Universidad de Alicante,
  Campus de San Vicente del Raspeig, E-03690 Alicante, Spain.
}
\affiliation{
  London Centre for Nanotechnology, University College London, 17-19 Gordon Street, WC1H0AH, London, United Kingdom
}
\author{D. Jacob}
\affiliation{
  Max-Planck-Institut f{\"u}r Mikrostrukturphysik, Weinberg 2, 06120 Halle, Germany
}
\author{C. Untiedt}
\affiliation{
  Departamento de F{\'i}sica Aplicada, Universidad de Alicante,
  Campus de San Vicente del Raspeig, E-03690 Alicante, Spain.
}
\email[Electronic mail address:]{untiedt@ua.es}

\date{\today}

\begin{abstract}
Atomic contacts made of ferromagnetic metals present zero-bias anomalies in the 
differential conductance due to the Kondo effect. 
These systems provide a unique  opportunity to perform a statistical analysis of 
the Kondo parameters in nanostructures since a large number of contacts can be 
easily fabricated using break-junction techniques. The details of the atomic structure 
differ from one contact to another so that a large number of different configurations 
can be statistically analyzed. Here we present such a statistical analysis of the Kondo effect 
in atomic contacts made from the ferromagnetic transition metals Ni, Co and Fe.
Our analysis shows clear differences between materials that can be understood by 
fundamental theoretical considerations. This combination of experiments and theory 
allows us to extract information about the origin and nature of the Kondo effect in 
these systems and to explore the influence of geometry and valence in 
the Kondo screening of atomic-sized nanostructures.
\end{abstract}

\maketitle

\section{Introduction}
\label{sec:intro}

The Kondo effect is one of the most intriguing phenomena arising from electronic correlations
which was first observed over 80 years ago as a then unexpected increase of the resistance of
gold wires at very low temperatures.\cite{deHaas_firstkondo} 
This phenomenon was successfully explained by Kondo 30 years later in his seminal 
work\cite{Kondo_original} as due to scattering of conduction electrons off magnetic 
impurities present in the Au samples, thereby screening its magnetic moment.
More generally, whenever a local magnetic moment is coupled to a sea of conduction electrons,
the Kondo effect can arise at low enough temperatures with important consequences for the
electronic and magnetic properties of the system.

Also in the case of mesoscopic devices, the Kondo effect strongly alters the 
electronic structure and therefore has dramatic consequences on the transport characteristics 
of the system. One of the simplest mesoscopic devices showing the Kondo effect is the case of a
quantum dot connected in series to two metallic 
leads:\cite{Goldhaber_KondoSET,Cronenwett_KondoQdot} 
When an electronic level of a quantum dot is well below the Fermi Energy of the metallic leads
and the Coulomb repulsion is strong enough to prevent double occupation, the  quantum dot
behaves as a magnetic impurity. In this situation conduction through the quantum dot is usually 
strongly suppressed due to the Coulomb blockade \cite{kastner_cbreview}. 
However, at low temperatures the Kondo effect restores the conductance due to the apparition of 
a sharp resonance --the so-called Kondo resonance-- in the spectral function of the quantum dot 
right at the Fermi energy of the electrodes.

The essence of the Kondo effect is the formation of a total spin-singlet state between 
the impurity electrons and the conduction electrons near the Fermi level 
\cite{Hewson} below a certain critical temperature 
characteristic of the system, the Kondo temperature. 
The formation of this Kondo singlet state gives rise to the effective screening of the magnetic 
moment of the impurity, and leads to the formation of a sharp resonance in the spectral 
density of the impurity electrons right at the Fermi level. This is the affore mentioned 
Kondo resonance, sometimes also called Abrikosov-Suhl resonance\cite{Abrikosov,Suhl,Nagaoka}.
In the case of magnetic impurities in metallic host materials, the formation of the Kondo
resonance in the spectral density of the impurity leads to additional scattering of the
conduction electrons, resulting in the increase of the resistance of the metal at low
temperatures (for a review of the Kondo effect in bulk metals with magnetic impurities 
see e.g. Ref. \onlinecite{Hewson}).

Other systems where evidence for the Kondo effect has been found by the manifestation of  
a Kondo resonance either in the spectral density or the conductance characteristics, include 
point contacts \cite{kondoqpc,Yanson_pointcontacts}, different molecules containing magnetic 
atoms on surfaces\cite{divanadium,molecco}, fullerenes\cite{csesentanormal}, carbon 
nanotubes contacted by metallic electrodes\cite{nanotubes,kim_fanont} and magnetic atoms on 
surfaces studied by Scanning Tunneling 
Microscopy (STM).\cite{Madhavan_science,Li_adatom,sander_adatom,Ternes_review,Madhavan_prb,
Knorr_adatomsdistance,Wahl_adatom,Nagaoka_temp,Jamneala_FeCoNi}

In the case of magnetic adatoms on metal surfaces studied by STM, interference of 
different conduction channels through the atom (one of them bearing the Kondo resonance), 
gives rise to Fano lineshapes \cite{Fano_original} in the low-bias conductance characteristics,
similar to the case of a quantum dot coupled laterally to a wire.\cite{gores_fano,sato_fano} 
By fitting those lineshapes to the Fano model one can obtain different parameters that 
describe the characteristics of the Kondo screening in the system, i.e. the width, position 
and amplitude of the Kondo peak (for a review see e.g. Ref. \onlinecite{Ternes_review}).
Of special relevance to our work are the STM experiments performed in the high conductance 
regime when the tip is brought into contact with the adatom\cite{neel_contact,vitali_contact,neel_cluster,kroger_review_2008,neel_silver,neel_quantized,kroger_heating,
bork_twoimpurities,Berndt_Fetip}. 

In a recent work we reported the observation of Kondo-Fano lineshapes in the conductance
characteristics of atomic contacts made from ferromagnetic materials \cite{Calvo_Nature}. 
In contrast to 

STM experiments where the contacted adatom can be imaged
and the 
geometry of the system formed by the adatom and the surface can be completely 
characterized, break junction experiments do not allow to control the geometry of 
the system to the same extend 
(although normally give rise to some geometrical repetition\cite{annealing}). On the other hand, using break junctions it is possible to form and study a 
large number of different configurations.\cite{Review_Contacts}

The observation of the Kondo effect in ferromagnetic atomic contacts was highly unexpected for two reasons: First,
given the chemical homogeneity of the atomic contacts, a devision into magnetic impurity and a 
spin-degenerate Fermi sea, as required for the appearance of the Kondo effect, is not obvious 
at all.
Second, electrodes and the contact atoms are made from a ferromagnetic material.
However, Kondo effect and ferromagnetism are generally competing phenomena: 
For example, a strong enough magnetic field corresponding to a Zeeman energy 
above the binding energy of the Kondo singlet (about the Kondo temperature)
will break up the singlet and thus the Kondo effect\cite{splitfieldtheory}
while, on the other hand, magnetic fields on the order of the Kondo temperature 
or below will lead to a splitting of the Kondo resonance as shown e.g. in Refs. 
\onlinecite{Cronenwett_KondoQdot, Kogan_splitting, Quay_splitting}.

Likewise the Kondo resonance splits when a quantum dot is connected to ferromagnetic 
electrodes in the case of parallel alignment of the two electrodes' magnetic polarizations 
while for antiparallel alignment of the electrodes a normal Kondo effect is obtained
\cite{Martinek_qdotferro,choi_qdotferro,Hauptmann_qdotferro,Pasupathy_c60ferro}. 
Therefore one would expect that the coupling of the contact region to the strongly 
ferromagnetic bulk electrodes should either eliminate the Kondo effect completely 
or at least split the Kondo resonance unless the antiferromagnetic coupling between 
magnetic impurity and conduction electrons is strong enough compared to any other 
interactions.

Here as in our previous work\cite{Calvo_Nature}, we propose that the Kondo effect in 
ferromagnetic atomic contacts originates from individual $d$-levels of the undercoordinated tip 
atoms of the nanocontact, and the Kondo screening is due to the delocalized $sp$-electrons which 
are basically spin-unpolarized. Hence the impurity or quantum dot bearing the spin consists of 
one or several $d$-levels of an individual contact atom. 

In our model, depending on the material the spin of the tip atom can be localized in different 
$d$-levels. Due to the low symmetry of the contact atoms the individual $d$-levels couple 
differently to the $sp$-conduction electron bath, resulting in different Kondo screenings of the 
spin. This scenario may explain the different Kondo behaviour observed for contacts made from 
Fe, Co and Ni.\cite{Calvo_Nature} 
Contacts made of the same material may also present slightly different atomic configurations 
which will also influence their Kondo properties. In our experiments, from the fitting of the 
differential conductance curves performed on atomic-sized contacts of ferromagnetic materials 
to a Fano expression, we extract the relevant parameters that characterize the Kondo effect 
on each contact. 

In this work, the validity of the Fano-Kondo model for the contact regime and more specifically 
for the case of atomic contacts is revised and discussed. We present a more exhaustive analysis 
than in Ref. \onlinecite{Calvo_Nature} for all the parameters of the Fano-Kondo model in our system and 
extract new information from the data, for instance, the distributions for the occupation of the $d$-level for Fe, Co 
and Ni.  This further analysis confirms the marked differences of the case of Ni with respect to Fe 
and Co. We propose here a new interpretation of the data: the nature of the Kondo screening is 
different for Ni contacts than for Fe and Co. From the combined statistical analysis of the Kondo 
parameters and theoretical fundamental considerations, we can deduce the influence of valence and 
environment on the Kondo screening in nanostructures.

This paper is organized as follows: The first two sections are introductory. First, we review 
the transport properties of atomic contacts in Sec. \ref{sec:transport}.
Then in Sec. \ref{sec:basic_theory}, we present the basic elements of the theory of the
Kondo effect in the framework of the Anderson impurity model and justify the validity of the Fano-Kondo model
for the case of atomic contacts of transition metals.
Secs. \ref{sec:expdetails} and \ref{sec:results} are devoted to the experimental methods and the 
presentation and discussion of the experimental results respectively. 
We finally discuss the results in the framework of the theory presented in Sec. 
\ref{sec:discussion} and summarize our main conclusions in Sec. \ref{sec:conclusions}.

\section{Transport in atomic contacts}
\label{sec:transport}

The system under study in this work are atomic contacts\cite{Review_Contacts}, i.e. a contact 
between two metallic leads mediated by an atom forming a metallic bond with both leads. More 
specifically, we focus here on the case of homogeneous contacts, those where all the atoms 
forming the structure, leads and contact, are of the same element.
The formation of an atomic contact can be identified from electronic transport measurements. 
When two pieces of the same pure metal are brought into contact, the conductance  plotted 
against the inter-electrode distance shows a plateau when the atomic contact is formed, close to 
the quantum of conductance $G_0 = 2e^2/h$. The exact value of this quantity will depend on the 
material and the atomic configuration of the contact. In general, the atomic orbitals will define
 a number of eigen-channels, each of these with a transmission probability $T_i$ which will be 
reflected in the total conductance through the Landauer formula $G = G_0 \sum_i T_i$. The number 
of channels will be related to the valence of the metal\cite{Elke_Nature,Cuevas_PRL98} and will normally 
include a high transmitive $s$-type channel and several other with lower transmission, resulting 
in a final conductance in the range of 0.7-2.5 $G_0$. Experimentally the evolution of the 
conductance can be recorded over the formation or breaking process of the contact showing
in most of the cases\cite{Untiedt_PRL07} 
plateaus, not only for the one-atom contact but for every atomic rearrangement of the wire 
while pulling. 

\begin{figure}[htp]
\centering
\includegraphics[width=\linewidth]{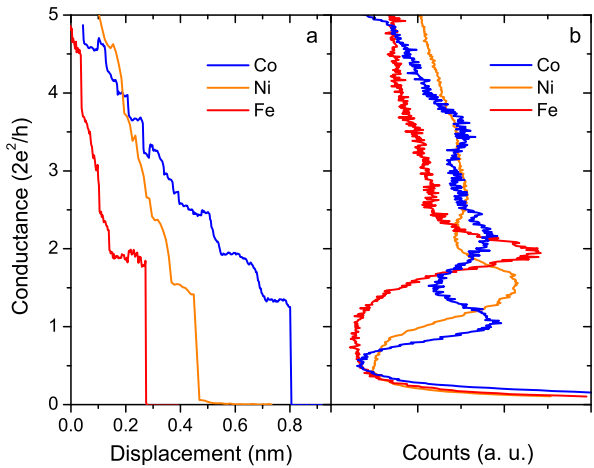}
\caption{\label{fig:traces}
  (a) Typical traces of conductance for breaking contacts of Co, Fe and Ni. 
  (b) Histograms of conductance are constructed from thousands of these breaking traces of 
  contacts fabricated by STM at 4.2 K for Ni, Fe or Co. The histograms for the three materials 
  show a clear first peak at values above $2e^2/h$.
}
\end{figure}

The one-atom plateau of conductance for the case of gold is near the quantum of conductance 
and, as shown in Fig. \ref{fig:traces}a, higher for the case of the $3d$-transition metals 
Fe, Co and Ni. From many of these traces we can build histograms which help us to identify the 
transport properties of the most probable atomic configurations of the contacts.

Different authors\cite{Untiedt_absence,Ludolph_fluctuations} agree that under cryogenic 
conditions and considering a high enough amount of data (thousands of conductance traces), the 
conductance of monatomic contacts made from Fe, Co or Ni takes a value higher than the quantum 
of conductance $G_{0}=2e^{2}/h$, as expected in general for transition 
metals.\cite{Review_Contacts}
Histograms of conductance of Fe, Co and Ni constructed from thousands of breaking traces at 
4.2~K are shown in Fig. \ref{fig:traces}b. The most probable values for the conductance of the 
monatomic contacts are 1.2~$G_{0}$, 1.6~$G_{0}$ and 2$G_{0}$ for Co, Ni and Fe, respectively.
Conductance values between 1~$G_0$ and 2~$G_0$ for ferromagnetic nanocontacts are in overall 
agreement with theoretical calculations.\cite{Jacob_Ni,Smogunov_Ni,Bagrets_CoNi,Haefner_FeCoNi}
The conduction of these atomic contacts is the sum of the contributions of different channels, 
where the $s$-channel is expected to be open and practically degenerate in spin and thus to have 
an associated conductance of nearly $2e^2/h$. 
In the ferromagnetic metals Fe, Co and Ni another contribution to the overall conductance
comes from the transmission of electrons via the five $3d$-orbitals. However, due to the 
directionality and stronger localization of the $d$-orbitals, the electrons in the $d$-channels 
are easily scattered. Hence the transmission in $d$-channels is usually far from perfect as has 
been shown by {\it ab initio} calculations\cite{Jacob_Ni}.
In this work it was also shown, that only the $d$-channels are spin-polarized while the 
$s$-channel basically is unpolarized.

\section{Basic elements of the theory}
\label{sec:basic_theory}

As discussed in the introduction, in the experiments the presence of the Kondo effect is 
reflected in the conductance of a one-atom contact as a zero-bias anomaly. The details of 
this anomaly can be related to the characteristics of the Kondo effect in the framework of 
the Anderson impurity model (AIM).\cite{Anderson} The purpose of this section is to provide 
the basic elements for a theoretical description of the Kondo effect in the framework of the 
single-level Anderson impurity model (1AIM) and the two-level Anderson impurity model (2AIM). 
This will allow us to properly analyse our experiments for different materials. 
Additionally, we develop a simple microscopic model in order to understand the occurrence of
different Fano lineshapes within the same material in terms of the variation of microscopic
interactions due to variation in the atomic structure in the contact region.

\subsection{Model of a nanocontact}

We assume that the Kondo effect takes place in the under coordinated tip atoms of the 
nanocontact. This assumption seems reasonable considering that the Kondo effect is not 
observed in bulk samples of the ferromagnetic materials studied here. Hence the Kondo 
effect must be related to the atomic-size constriction of the nanocontact. 
Furthermore, the Coulomb interaction and the localization of the electrons within the 
atomic-size constriction, and especially at the tip atoms should be enhanced as compared 
to bulk. 

\begin{figure}
  \begin{center}
    \includegraphics[width=0.9\linewidth]{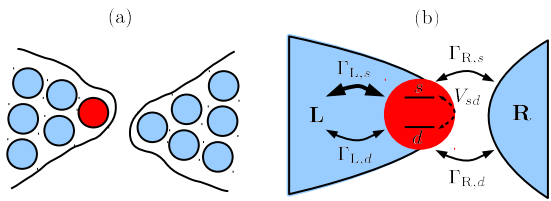} 
  \end{center}
  \caption{
    (a) Cartoon of nanocontact just before breaking. 
    The tip atom (red) of the left electrode is only 
    weakly coupled to the right part of the nanocontact
    (and vice versa). This leads to the simplified model
    of the nanocontact as shown in (b). In the simplified
    model the tip atom (red) consists of a non-interacting
    $s$-level and a strongly interacting $d$-level
    which couple more strongly to the left electrode
    than to the right electrode ($\Gamma_{\rm L,\alpha}>\Gamma_{\rm R,\alpha}$).
    The hybridization $V_{sd}$ between $s$- and $d$-level
    is due to crystal field splitting and is weak compared 
    to the other energies. Also the coupling of the 
    $d$-level to the electrodes is much weaker than
    the coupling of the $s$-level ($\Gamma_s\gg\Gamma_d$).
    \label{fig:model}
  }
\end{figure}

Typically, the nanocontacts of the ferromagnetic metals Fe, Co and Ni form dimers in the last 
step before breaking\cite{Pauly_NiHisto,Calvo_ieee} as illustrated by the cartoon in 
Fig. \ref{fig:model}(a). 
In his case each tip atom will be more strongly coupled to one electrode than to the other.
As was said before in Sec. \ref{sec:transport} the main conduction channel through the tip 
atoms is the spin-degenerate $s$-channel with nearly perfect transmission. The $d$-channels
on the other hand are strongly spin-polarized and their transmission is weaker or even 
completely blocked due to scattering by the geometry.\cite{Jacob_Ni} 
As we have shown in our previous work\cite{Calvo_Nature} by means of {\it ab initio} calculations,
due to disorder and low coordination in the contact region one of the $d$-levels of the tip 
atom couples only very weakly to the $d$-levels of the neighbouring atoms, and instead only 
couples to the basically spin-degenerate $s$-channel. 
Now this is the situation where the Kondo 
effect can take place and which is described by the Anderson impurity model: A strongly 
interacting $d$-level couples to a non-interacting sea of conduction electrons. Hence we can 
identify this $d$-level with the ``quatum dot'' or impurity in our experiments. This situation 
is schematically depicted in Fig. \ref{fig:model}(b).

This scenario is further supported by recent Dynamical Mean-Field Theory calculations of a 
Ni nanocontact connected to electrodes made from Cu instead of Ni, thus neglecting the 
ferromagnetic coupling to the bulk electrodes.\cite{Jacob_DMFT}
In this situation a Kondo effect emerges in one of the $d$-channels of the tip atoms of the 
Ni nanocontact with the Kondo temperature in good agreement with the ones measured for Ni 
nanocontacts. 

\subsection{Kondo effect in the Anderson model}

As we have argued in our previous work on the basis of {\it ab initio} calculations
the combined effect of under coordinated tip atoms and disorder in the contact region 
can lead to the selection of a single $d$-level of the tip atom that is only weakly coupled
to the $d$-levels of the neighbouring atoms. Hence the magnetic coupling to the 
neighbouring atoms is reduced for this level and therefore the Kondo effect can arise
if this level is half-filled. This situation can be fulfilled in the case of Ni nanoconctacts 
where the $d$-shell is $d^9$ and thus has a hole. In this case the single-level 
Anderson impurity model (1AIM)
where a single strongly interacting $d$-level is coupled to a bath of non-interacting 
conduction electrons is a good description of the situation:
\begin{eqnarray}
  \label{eq:AIM}
  \hat\mathcal{H}_{\rm AIM}&=&\epsilon_d\,\hat{n}_d + U \hat{n}_{d\uparrow}\hat{n}_{d\downarrow}
  + \sum_{q,\sigma} \epsilon_q \hat{c}_{q\sigma}^\dagger \hat{c}_{q\sigma}
  \nonumber\\
  && +\sum_{q,\sigma} (V_q \hat{c}_{q\sigma}^\dagger \hat{d}_\sigma
  + V_q^\ast \hat{d}^\dagger_\sigma \hat{c}_{q\sigma})
\end{eqnarray}
where $\epsilon_d$ is the energy of the $d$-level, $U$ is the (effective) Coulomb
repulsion between two electrons in the $d$-level, $\epsilon_q$ is the energy dispersion
of the bath electrons, and $V_q$ is the coupling (or hopping) between the bath and
the $d$-level. 
In the case of Fe and Co there is more than one hole in the $d$-shell and hence one 
should consider a multi-level AIM. We will do so in the next subsection.

At zero temperature the Anderson model is a Fermi liquid and hence has a 
(renormalized) quasi-particle resonance (i.e. the Kondo peak in the Kondo 
regime) near the Fermi level. The Green's function of the $d$-level is given by:
\begin{equation}
  \label{eq:Gd}
  G_d(\omega) = \frac{z}{\omega-\epsilon_K+i\Gamma_K}
\end{equation}
where $z$ is the quasi-particle weight (i.e. the renormalization of the 
single-particle wave-function due to many-body effects), $\epsilon_K$ is 
the position of the quasi-particle peak with respect to the Fermi level 
(set to zero for convenience), and $\Gamma_K=k\,T_K$ is half the width of 
the quasi particle resonance which defines the Kondo Temperature. 
Correspondingly, the projected density of state of the $d$-level is 
a Lorentzian centred at $\epsilon_K$ and width $2\Gamma_K$:
\begin{equation}
  \label{eq:rhod}
  \rho_d(\omega) = -\frac{1}{\pi}{\rm Im}\,G_d(\omega) = \frac{z\,\Gamma_K/\pi}{(\omega-\epsilon_K)^2+\Gamma_K^2}
\end{equation}

In the case of the single-level AIM, the Kondo temperature $T_K$ can be easily calculated 
from the parameters of the model as follows:\cite{Hewson}
\begin{eqnarray}
\label{eq:tk}
\Gamma_K=kT_{K} &=& \frac{\sqrt{\Gamma_dU}}{2}e^{\pi \epsilon_{d}(\epsilon_{d} +U)/\Gamma_d U}
\end{eqnarray}
where $\Gamma_d$ is the broadening of the $d$-level due to the coupling to the 
bath obtained by integrating out the bath degrees of Freedom:
\begin{equation}
  \Gamma_d = V^2 \rho_{\rm bath}(\omega=0)
\end{equation}
Here we have assumed an approximately constant bath density
of states $\rho_{\rm bath}$ and the coupling $V$ independent of $q$.
Note the exponential dependence of the Kondo temperature 
on the interaction $U$ and broadening $\Gamma_d$. This 
means that mild changes in the parameters can have a huge
effect on the Kondo temperature.
Also note that other definitions of the Kondo temperature 
may differ by a constant prefactor.

Since the 1AIM is a Fermi liquid at zero temperature\cite{Nozieres}
we can exploit further relations of the Fermi liquid theory.
For example, one obtains the following important relationship
between the impurity-level occupation and the Kondo parameters
(see e.g. Ch. 5 in Ref. \onlinecite{Hewson}): 
\begin{equation}
  \label{eq:1AIM-occup}
  n_d = 1 - \frac{2}{\pi}\arctan\left(\frac{\epsilon_K}{k\,T_K}\right)
\end{equation}
Also from the Fermi liquid theory of the Anderson model we obtain the 
following exact relation between the density of states at the Fermi level 
$\epsilon_F\equiv 0$ (in general not the maximum of the Kondo peak) 
to the occupation of the $d$-level and the broadening $\Gamma_d$ due to 
the coupling of the $d$-level to the rest of the system:
\begin{equation}
  \label{eq:1AIM-rhod0}
  \rho_d(0) = \frac{\sin^2(\frac{\pi}{2}n_d)}{\pi\Gamma_d}
\end{equation}
Finally the \emph{amplitude} of the Kondo resonance is
\begin{equation}
  \label{eq:1AIM-amplitude}
  A_K\equiv\rho_d(\epsilon_K) = \frac{z}{\pi k\,T_K}
\end{equation}

\subsection{Underscreened Kondo effect in the multi-level Anderson model}
\label{sec:UKE}

Co and Fe feature 2 and 3 holes, respectively, in the $3d$-shell of each atom and hence have 
an atomic spin $S>1/2$ due to Hund's rule coupling. Therefore a description in terms of a 
1AIM as before is problematic. Nevertheless the experimental results can be fitted 
very well to a 1AIM (see below and Sec. \ref{sec:results}).
The explanation might be that we are really dealing with a so-called underscreened Kondo 
effect\cite{Nozieres_RealKondo} (UKE) where only a spin-$1/2$ in one of the $d$-levels is 
screened while the rest remains unscreened. 

Such an UKE behaves in many ways like a normal (fully screened) $S=1/2$ Kondo effect. For example, 
it is still characterized by a zero-bias anomaly resulting from a sharp resonance in the screened 
impurity level. Further support for this hypothesis comes from a paper by Perkins {\it et al.}:\cite{Perkins_EPL} 
They find that for an underscreened Kondo lattice, ferromagnetism and Kondo effect can in fact coexist. 
On the other hand, the UKE has a number of peculiar consequences such
as the formation of a so-called singular Fermi liquid state characterized e.g. by the 
divergence of the quasi-particle weight and thermodynamic quantities such as the specific 
heat capacity\cite{Coleman_PRB03,Coleman_PRB05}.
Note that in this sense the UKE is very similar to the so-called ferromagnetic Kondo effect\cite{Koller_fmkondo} 
where the impurity spin couples ferromagnetically with the conduction electrons. The resulting antiscreening 
of the magnetic moment by the conduction electrons also leads to the formation of a singular Fermi liquid 
state. The ferromagnetic Kondo effect has recently been discussed theoretically in the context of magnetic 
impurities in nanocontacts.\cite{Gentile_fmkondo,Lucignano_natmat}

\begin{figure}
  \begin{center}
    \includegraphics[width=0.8\linewidth]{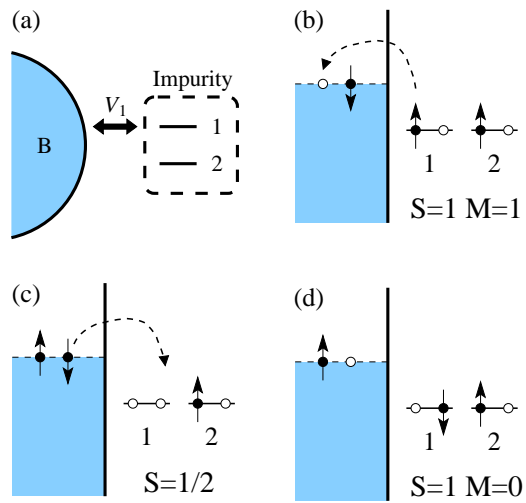} 
  \end{center}
  \caption{\label{fig:UKE}
    (a) Schematic drawing of 2-level Anderson impurity model in the underscreened situation.
    Only level 1 of the impurity is coupled to the bath B.
    (b-d) Schematic illustration of a hopping process contributing to the underscreened Kondo effect
    in the 2-level Anderson impurity model.}
\end{figure}
 
We consider a two-level Anderson impurity model (2AIM). This should model the situation of 
Co which has two holes in the $d$-shell. As in the case of the simple AIM, the system is 
divided into two subsystems: the conduction electron bath B and the impurity I with the two 
interacting levels. The version of the 2AIM model that is relevant for the underscreened Kondo 
effect is depicted schematically in Fig. \ref{fig:UKE}(a): Only one of the 
impurity levels couples to the conduction electron bath while the coupling of the other level 
is negligible.\cite{Posazhennikova_PRB,Logan_PRB}
Hence the Hamiltonian of the 2AIM is given by: 
\begin{eqnarray}
  \label{eq:2AIM}
  \lefteqn{
  \hat{H}_{\rm 2AIM} = \hat{H}_{\rm I} + \hat{H}_{\rm B} + \hat{H}_{\rm T}} \nonumber\\
  &=& \sum_i \left( \epsilon_i \hat{n}_i + U\,\hat{n}_{i\uparrow} \hat{n}_{i\downarrow} \right) 
  + U^\prime\, \hat{n}_1 \hat{n}_2 - J_{\rm H}\, \hat{\mathbf{s}}_1\cdot\hat{\mathbf{s}}_2
  \nonumber\\
  &&+\sum_{q,\sigma} \epsilon_q c_{q\sigma}^\dagger c_{q\sigma}
  + \sum_{q,\sigma} V_{1,q} \left( d_{1\sigma}^\dagger c_{q\sigma}  + c_{q\sigma}^\dagger d_{1\sigma} \right)
\end{eqnarray}
where $d_{i\sigma}$ ($d_{i\sigma}^\dagger$) destroys (creates) one electron in impurity level $i$ 
with spin $\sigma$, and $\hat{n}_{i\sigma}=d_{i\sigma}^\dagger d_{i\sigma}$ is the occupation number 
operator for level $i$ and spin $\sigma$ and $\hat{n}_i=\hat{n}_{i\uparrow}+\hat{n}_{i\downarrow}$. 
$\hat{\mathbf{s}}_i=\sum_{\sigma\sigma^\prime} d_{i\sigma}^\dagger \vec\tau_{\sigma\sigma^\prime} d_{i\sigma^\prime}$ 
measures the spin in level $i$.
$\epsilon_i$ are the energies of the two impurity levels, $U$ is the Coulomb repulsion within
the same level $i$ and $U^\prime$ is the Coulomb repulsion between electrons in different levels
which is generally smaller than $U$, and $J_{\rm H}$ is the Hund's rule coupling. 
The bath B is described as in the case of the 1AIM, eq. (\ref{eq:AIM}).
Finally, only impurity level 1 is coupled to the conduction 
electron bath with hopping $V_1$, while the coupling of impurity level 2 is negligible.

For the sake of simplicity we also assume that the two impurity levels are degenerate:
$\epsilon_1=\epsilon_2=\epsilon$. 
Typically, the \emph{intra}-level Coulomb repulsion $U$ is bigger than the \emph{inter}-level
Coulomb repulsion $U^\prime$ by an amount of the order of the Hund's rule coupling: 
$U \approx U^\prime + J_{\rm H}$.
Assuming a constant bath density of states and $q$-independent coupling $V_{1,q}=V_1$,
the half-width of level 1 due to the coupling to the conduction electrons is given by 
$\Gamma_1={V_1}^2\cdot\rho_{\rm bath}$. 

Now in the situation where the two impurity levels are well below the Fermi energy of the
conduction electrons ($\epsilon_F > 2\epsilon+U^\prime$), and the {\it intra}-level Coulomb 
repulsion $U$ is strong enough to prevent double occupation of each impurity level 
($2\epsilon+U > \epsilon_F$) the impurity will be doubly occupied.
And due to Hund's rule coupling the impurity will then be in a total spin-triplet state, i.e. 
will have total spin $S_{\rm I}=1$ (See App. \ref{app:2AIM} for further details). 

In this situation, switching on the coupling $H_{\rm T}$ between the impurity and
the conduction electron bath gives rise to hopping processes as depicted schematically
in Fig. \ref{fig:UKE} which will {\it partially} screen the total spin-1 of the impurity
by flipping the spin in impurity level 1. This partial screening of the impurity spin $S>1/2$  
by a single conduction electron channel is called {\it underscreened Kondo effect} 
(UKE).\cite{Nozieres_RealKondo} 
The coupling to the residual spin in the other impurity level gives rise to a 
so-called {\it singular Fermi liquid} (SFL) behaviour\cite{Coleman_PRB03,Posazhennikova_PRL,Mehta_PRB}, 
in contrast to the normal Fermi liquid behaviour of the usual fully screened Kondo effect.

The SFL is characterized by a cusp in the spectral density at low temperatures, i.e
for low energies the spectral density of impurity level 1 is approximately given by:
\cite{Koller_PRB,Logan_PRB}
\begin{equation}
  \rho_1(\omega) \approx \frac{1}{\pi\Gamma_1}\left(1-\frac{b}{\ln(|\omega|/kT_0)^2}\right)
\end{equation}
where $T_0$ is a new temperature scale associated with the Spin-1 UKE
and $b>0$ is a constant.
The cusp in the spectral density is related to a logarithmic divergence 
of the quasi-particle weight in the underscreened Kondo regime:\cite{Coleman_PRB05}
$z\propto1/\omega\log(kT_0/\omega)$.
Hence in contrast to the normal Kondo effect there is no well defined
quasi-particle associated with the UKE. Hence the name \emph{singular} 
Fermi liquid. Note that although the quasi-particle weight $z$ diverges for 
$\omega\rightarrow 0$, the spectral density $\rho_1(\omega)$ itself 
does not diverge. 

In an actual experiment the logarithmic cusp characteristic for the UKE is
probably hard to resolve due to limited resolution and the smoothening effect
of finite temperature. Hence in practice the zero-bias anomaly arising from the
UKE is undistiguishable from that arising from the normal Lorentzian-type Kondo 
peak unless very low temperatures can be reached and the experimental resolution 
is fine enough to resolve the cusp.

The half width of the resulting UKE resonance is determined by the temperature 
$T_0$ as: $\Gamma_{K}^{S=1}=kT_0\exp(-\sqrt{2b})$.
In order to compare this width with the one of the normal Kondo peak in 
the 1AIM we consider the particle-hole symmetric regime. Following 
Ref. \onlinecite{Logan_PRB} we have 
$kT_0\propto \exp\left(-\frac{\pi}{4\Gamma_1}(U+J_{\rm H}/2)\right)$
while $kT_K\propto\exp(-(\pi U)/(4\Gamma_d))$.
Hence for the same parameters $U$ and $\Gamma_d=\Gamma_1$, the width of the 
resonance should be smaller in the case of the UKE than for the normal Kondo 
effect due to Hund's rule coupling $J_{\rm H}$ and the additional exponential 
factor $\exp(-\sqrt{2b})<1$. Assuming that the interaction $U$ and coupling 
$\Gamma$ for the $d$-level giving rise to the Kondo effect 
is approximately the same for the three transition metals considered 
here, this might explain why the widths of the zero-bias anomalies obtained for the 
cases of Fe and Co are considerably smaller than for Ni (see Sec. \ref{sec:results}).

Finally, for the 2AIM in the UKE regime one obtains exactly the same formula relating 
the occupation $n_1$ of the impurity level 1 (the one coupled to the conduction electron 
bath) to the width and position of the resonance as in the case of the 1AIM, 
eq. (\ref{eq:1AIM-occup}).\cite{Logan_PRB} This explains why the results in the case of Co 
and Fe can be fitted so well to the formula for the 1AIM (see Sec. \ref{sec:results}).

\subsection{Kondo-Fano lineshapes}

The goal of this section is to devise a simple model in order to understand the 
occurrence of different Fano lineshaps (i.e. peaks, dips or asymmetric Fano curves) 
for the same material.  
Our simple model also demonstrates the complicated dependence of the Fano parameters 
on the basic microscopic parameters. Fano lineshaps in a related model have recently 
been studied by Zitko\cite{zitko_fano} using the Numerical Renormalization Group and 
focusing on the temperature dependence of the Fano lineshapes. Here we neglect any 
temperature effects since the measured Kondo scales are much higher than the temperature 
of 4.2~K at which the experiments where performed. In any case the goal of this section 
is not to give an exact description of the conductance spectra but rather to obtain a
qualitative understanding of how different lineshapes emerge and how they depend on the 
microscopic details of the system.

\begin{figure}[htp]
  \centering
  \includegraphics[width=0.8\linewidth]{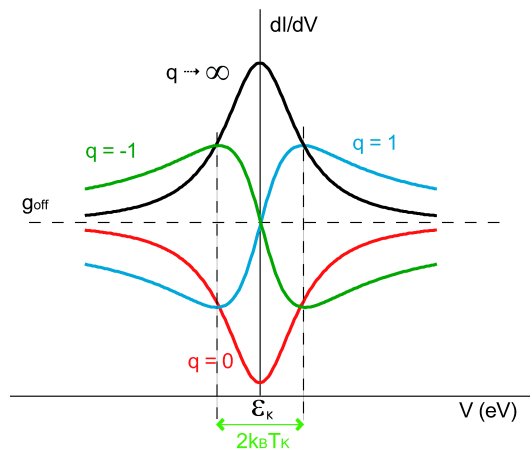}
  \caption{Graphical definition of the Fano resonance parameters.}\label{fig:fanopar}
\end{figure}

In the following we will show that the low-bias conductance ($\|eV\|\le\Gamma_K$) of our
simplified model of a nanocontact shown in Fig. \ref{fig:model} is well described by 
the Fano formula:
\begin{equation}
  \label{eq:Fano}
  G(V) = g_{\rm off} + \frac{A}{1+q^2}\,\frac{(\epsilon+q)^2}{\epsilon^2+1}  
  \hspace{1ex}\mbox{with}\,\epsilon = \frac{eV-\epsilon_K}{\Gamma_K}
\end{equation}
where $g_{\rm off}$ is the conductance offset, $A$ is the amplitude of the Fano resonance and 
$q$ is the Fano factor determining the shape of the Fano resonance (see Fig. \ref{fig:fanopar}),
and as described before $\epsilon_K$ is the energy position and $\Gamma_K$ the half-width
of the Kondo peak in the $d$-level which thus determine the position and width of the
resulting Fano lineshape.

As can be seen from Fig. \ref{fig:fanopar} when the Fano factor becomes very large 
($q\rightarrow\infty$) the conductance has a Lorentzian lineshape. This is the case when
the coupling $V_{sd}$ between the $s$- and the $d$-level of our tip atom is negligible so that 
the zero bias anomaly in the conductance is dominated by the direct transmission through the 
Kondo resonance of the $d$-level. On the other hand for $q=0$ we obtain a dip-feature in the 
conductance. This is for example the case when the direct transmission through the $d$-channel 
becomes negligible (e.g. for $\Gamma_{R,d}\approx0$) so that the conductance is only given by the
$s$-channel which features a Lorentzian dip due to the coupling $V_{sd}$ to the Kondo 
resonance in the $d$-level. For $|q|=1$ the Fano formula gives the typical asymmetric 
lineshapes. In this case the conductance is also dominated by the $s$-channel coupled
to the Kondo resonance in the $d$-level as for $q=0$ but now the $s$-level of the tip
atom is not near the Fermi level.

In the appendix we give a derivation of the Fano formula (\ref{eq:Fano}) for our simplified 
model of a nanocontact shown in Fig. \ref{fig:model}: The (left) tip atom where the Kondo 
effect is taking place is modeled by one $s$- and one $d$-level. While the $s$-level couples 
well to both electrodes (via $\Gamma_{L,s}$ and $\Gamma_{R,s}$) and thus has a nearly perfect 
transmission, the $d$-level hosting the Kondo resonance has a much weaker coupling to both 
electrodes ($\Gamma_{L,d}$ and $\Gamma_{R,d}$)). Generally we assume that the couplings to the 
left electrode are stronger than to the right electrode, i.e. $\Gamma_{L,\alpha}>\Gamma_{R,\alpha}$. 
Additionaly there is a small hybridization $V_{sd}$ between the $s$- and the $d$-level due to 
the crystal field.

As Meir and Wingreen showed in their landmark paper, at zero temperature and in linear
response the conductance through a nanoscopic conductor is well described by the Landauer 
formula even in the case of a strongly interacting system.\cite{Meir-Wingreen}
The conductance $G$ for small bias $V$ is then given in terms of the quantum mechanical 
transmission function $T(\omega)$ as $G(V)=G_0\times{}T(eV)$ where $G_0=2e^2/h$ is the 
fundamental conductance quantum.

As is shown in the appendix, the transmission $T(\omega)$ through the tip atom can 
be decomposed into the contributions of direct transmission through the indivdiual $s$- 
and $d$-channel, namely $T_s(\omega)$ and $T_d(\omega)$, respectively, and a mixed 
channel involving hopping between both channels, $T_{sd}(\omega)$:
\begin{equation}
  \label{eq:transm}
  T(\omega) = T_s(\omega) + T_d(\omega) + T_{sd}(\omega)
\end{equation}
Following eqs. (\ref{eq:Ts},\ref{eq:Td})
the direct channel transmissions $T_s$ and $T_d$ are given by the spectral densities
of the $s$- and $d$-level, respectively, and the couplings of the $s$- and $d$-levels to both 
electrodes $L$ and $R$.

The Kondo effect gives rise to the appearance of a Kondo resonance in the the $d$-level 
spectral function $\rho_d(\omega)$ given by eq. (\ref{eq:rhod}). Hence the contribution 
of the $d$-channel to the total transmission has a Lorentzian lineshape:
\begin{equation}
  \label{eq:dtransm}
  T_d(\omega) = \frac{z^2\,\Gamma_{L,d}\cdot\Gamma_{R,d}}{(\omega-\epsilon_K)^2+\Gamma_K^2}
  = \frac{4\,\Gamma_{L,d}\cdot\Gamma_{R,d}}{\Gamma_{L,d}+\Gamma_{R,d}}\cdot\frac{1}{1+x^2}
\end{equation}
where we have used $\Gamma_K=z(\Gamma_{L,d}+\Gamma_{R,d})/2$ and
we have defined the dimensionless quantity $x=(\omega-\epsilon_K)/\Gamma_K$.
Hence as pointed out above, the $d$-channel contribution to the transmission 
can only give rise to Fano-lineshapes with $q\rightarrow\infty$.

We assume that the unperturbed $s$-channel (i.e. without coupling to the $d$-level) 
has a featureless (i.e. flat) and almost perfect transmission $T_s^0\approx1$. Due 
to the coupling $V_{sd}$ to the $d$-level this transmission is modified according to 
(see appendix):
\begin{equation}
  \label{eq:stransm}
  T_s(\omega) = T_s^0 \cdot \left[ 1 + \frac{2\,z\,V_{sd}^2}{\Gamma_s\cdot\Gamma_K} \cdot
  \frac{(x+q_0)^2 -(x^2+1)}{(x^2+1)\cdot(1+q_0^2)} \right]
\end{equation}
where $\Gamma_s=\Gamma_{L,s}+\Gamma_{R,s}$ is the total broadening of the $s$-level and 
the dimensionless quantity $q_0$ has been defined as the ratio between the $s$-level energy
and the $s$-level broadening: $q_0=-2\epsilon_s/\Gamma_s$ (see appendix for details).
The second term of the r.h.s. is of the Fano form, eq. (\ref{eq:Fano}) and represents
the modulation of the almost perfectly transmitting $s$-channel due to the coupling to the
Kondo resonance in the $d$-channel.
Since generally $2|\epsilon_s| < \Gamma_s$ we should have $|q_0|\le1$ and therefore 
the $s$-channel contribution to the transmission (\ref{eq:stransm}) can only
give rise to dip-like ($q_0\approx0$) or asymmetric Fano lineshapes $q\approx1$ but not
to the peak lineshapes where $|q_0|\gg1$.

For the mixed-channel contribution $T_{sd}$ to the total transmission we find the following
expression in the appendix:
\begin{equation}
  \label{eq:sdtransm}
  T_{sd}(\omega) = T^0_s \cdot \frac{z^2\,V_{sd}^2}{\Gamma_K^2} \cdot 
  \left( 
  \frac{\Gamma_{R,d}}{\Gamma_{R,s}} + \frac{\Gamma_{L,d}}{\Gamma_{L,s}}
  \right) \cdot \frac{1}{1+x^2}
\end{equation}
This contribution describes transmission processes where an electron hops from one
electrode to the $s$-level of the tip atom, subsequently to the $d$-level via $V_{sd}$, 
and then to the other electrode. Due to the Kondo peak in the $d$-level it gives rise 
to a Lorentzian lineshape in the transmission.

Hence we have shown that our model can give rise to all possible Fano lineshapes as 
obtained in the experiments. More specifically, the $s$-channel contribution 
$T_s(\omega)$ can give rise to the dip-like features ($q\approx0$) and the asymmetric 
Fano features ($|q|\approx1$) while the $d$-channel contribution $T_d(\omega)$
and the mixed channel contribution $T_{sd}(\omega)$ give rise to Lorentzian lineshapes
($|q|\rightarrow\infty$) in the transmission. Which term dominates depends on the 
specific amplitudes of the different transmission channels given by the basic 
parameters of our model.

It is of course possible to achieve the ``canonical'' Fano lineshape form for the 
conductance as in eq. (\ref{eq:Fano}) by summing up all the individual contributions 
to the total transmission (\ref{eq:transm}) and reorganizing the terms. 
We then obtain a ``new'' Fano factor $q$ different from the Fano factor $q_0$ for the 
pure $s$-channel contribution (\ref{eq:stransm}). This new Fano factor will depend on
$q_0$ and the amplitudes of the individual contributions to the transmission, and yields
a relatively complicated expression in terms of the basic parameters of our model.
The same is true for the amplitude $A$ of the Fano feature defined by eq. (\ref{eq:Fano}).

However, we can obtain quite simple expressions for $q$ and $A$ in eq. (\ref{eq:Fano})
in an important limit of our model, namely when the coupling of the $d$-level giving rise 
to the Kondo peak to one of the electrodes becomes very small, e.g. $\Gamma_{R,d}\rightarrow0$. 
In that case the direct transmission through the $d$-channel is strongly suppressed, i.e. 
$T_d\approx0$ so that now only the mixed channel contribution $T_{sd}$ can give rise to a 
Lorentzian lineshape in the transmission. 
\begin{eqnarray}
  \label{eq:fano-q}
  q &=& \sqrt{\frac{\Gamma_s}{\Gamma_{L,s}}}\cdot q_0 = 
  -\frac{2\epsilon_s}{\sqrt{\Gamma_s\cdot\Gamma_{L,s}}} \\
  \label{eq:fano-A}
  A &=& \frac{1+\Gamma_s/\Gamma_{L,s}}{1+q_0^2}\cdot
  \frac{4\,V_{sd}^2}{\Gamma_s\cdot\Gamma_{L,d}} \nonumber\\
  &=& 2\pi\cdot\frac{\Gamma_s}{\Gamma_{L,s}}\cdot
  \frac{\Gamma_{L,s}+\Gamma_s}{\Gamma_s^2+4\epsilon_s^2}\cdot{V_{sd}^2}\cdot{A_K}
\end{eqnarray}
where $A_K$ is the amplitude of the Kondo resonance in the spectral function of the $d$-level 
as given by eq. (\ref{eq:1AIM-amplitude}). Note that the Fano factor $q_0$ of the $s$-channel 
transmission $T_s$ is now scaled by the factor $\sqrt{\Gamma_s/\Gamma_{L,s}}>1$ to yield the 
Fano factor $q$ of the Fano lineshape in eq. (\ref{eq:Fano}) meaning that for large values of 
the ratio $\sqrt{\Gamma_s/\Gamma_{L,s}}$ we can obtain $|q|$ values $>1$ for $0<|q_0|\le1$. 
Furthermore, we see that the amplitude of the Fano resoance in the conductance is proportional 
to the amplitude of the Kondo resonance and to the square of the coupling $V_{sd}$ between the 
$s$-level and the $d$-level of the tip atom.

\section{Experimental details}
\label{sec:expdetails}

The experiments were performed using a home-made Scanning Tunnelling Microscope (STM) operated 
in a He cryostat at 4.2K. Two pieces of the same metal wire (Fe, Co or Ni) of 0.1 mm of diameter 
were scratched and sonicated in acetone and isopropanol before being mounted as 'tip and sample' 
in the microscope. 
The conductance between the two pieces of metal is obtained in a two 
terminal configuration by measuring the current at a fixed bias voltage, 
in this case 100 mV. In these conditions we can record traces of conductance wile changing 
the distance between the two metals (a typical trace is shown in Fig. \ref{fig:traces}a) 
in a similar fashion as performed in other break junction experiments \cite{Review_Contacts}. 
The samples are then prepared at low temperatures by indentation until no traces with 
subquantum events are shown.
The histograms (Fig. \ref{fig:traces}b) are similar to those at Mechanically Controlled
Break Junction (MCBJ) experiments where a fresh surface is formed at cryogenic vacuum when 
breaking a notched wire by the controlled bending of a elastic substrate (See e.g. Ref. 
\onlinecite{Review_Contacts} for details). This shows that our measurements are performed 
over a clean spot of our samples. A strong indentation between the two electrodes is performed 
between the fabrication of consecutive atomic contacts to ensure the cleanliness. 
Our STM set-up has imaging capabilities, however, our surfaces are not atomically flat due
to preparation and 
in order to acquire a large number of contacts to analyse
no imaging is performed between contacts formation.

As described in Sec \ref{sec:transport}, traces show plateaus coming from the atomic 
rearrangement of the wire while pulling and the last plateau around the quantum of 
conductance is associated with the formation of a  single atom contact. 
Once the conductance of the one-atom contact has been determined by the position of 
the first peak of the conductance histogram, we can fabricate monoatomic contacts by 
stopping the breaking process of the contact at the desired value of conductance within 
the range defined by the conductance histogram. The stability of our system allows us to 
maintain such an atomic contact for hours.
We study the transport spectroscopy of these contacts in a similar way as performed in other
transport experiments either in tunnelling or high conductance regimes (e.g. the 
case of quantum dot devices or the spectroscopy of adatoms by STM). We sweep the 
bias voltage from -100 to 100 mV while recording the $dI/dV$ signal with the help of a 
lock-in amplifier when adding a 1mV AC excitation at a frequency about 1KHz to 
the applied bias voltage.

\section{Results}
\label{sec:results}

\begin{figure}[htp]
  \centering
  \includegraphics[width=\linewidth]{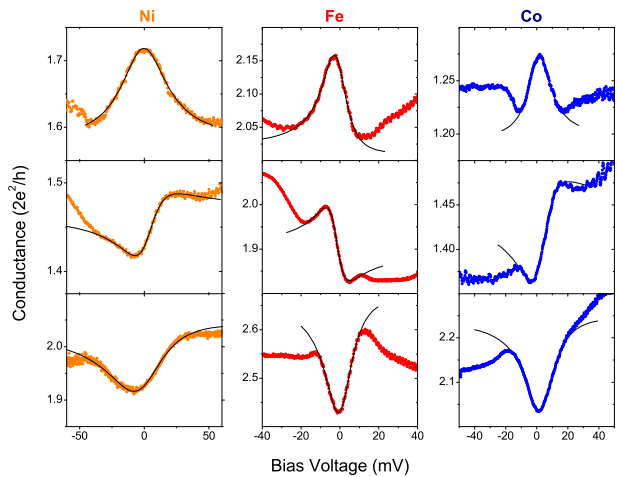}
  \caption{\label{fig:spectroscopy}
    Representative differential conductance curves for atomic contacts of Fe, Co and Ni. 
    A characteristic resonance associated with the Kondo effect appears at small bias. The 
    corresponding fit to the Fano lineshape is shown in black for each of these resonances. 
Due to the large differences in the width of the resonances, the fitting range is chosen so that is possible to 
fit the whole range of the resonance to the Fano lineshape.
    The selected curves exemplify different symmetry cases, from the more symmetric ($q\simeq\infty$ 
    and $q\simeq 0$ respectively) to the clearly asymmetric ones ($q\simeq \pm 1$). While all 
    possible symmetries seem to happen for the three materials, the width of the resonance 
    which is associated with the Kondo temperature, is different for the three materials, 
    especially for the case of Ni. 
  }
\end{figure}

The fabrication of atomic contacts by using a STM (or MCBJ) offers the possibility of studying a high number 
of different contact configurations in a reasonable amount of time. We fabricated hundreds of 
atomic contacts of Fe, Co and Ni and performed electron spectroscopy measurements as described 
in Section \ref{sec:expdetails}. About 80\% of these curves showed clear asymmetric profiles 
centred at zero bias as the ones plotted in Fig. \ref{fig:spectroscopy}.

Similar asymmetric zero bias anomalies (ZBA) were reported for ferromagnetic atomic contacts and 
attributed to the existence of conductance fluctuations \cite{ralph_rearrangements} or to the 
existence of a magnetic domain wall.\cite{Sekiguchi_zba_prb} On the other hand, these asymmetric 
profiles resemble the data reported for single magnetic adatoms in the contact regime, not only 
in shape but also in the energy scale of the 
features \cite{neel_contact,vitali_contact}.

In our previous work we have shown clear evidence of the Kondo effect 
being responsible for these ZBA.\cite{Calvo_Nature} Recently, ZBA profiles, possibly of similar 
origin, have been reported in Refs. \onlinecite{gloos1} and \onlinecite{gloos2} for nanocontacts 
made from other materials and also show the Kondo related ZBA in adatoms contacted by ferromagnetic tips
\cite{bork_twoimpurities,Berndt_Fetip}.

As in other Kondo systems, the asymmetric lineshapes can be fitted to the Fano equation 
(\ref{eq:Fano}). From this fitting we extract the values for the different parameters that 
describe the Kondo effect. As sketched in Fig. \ref{fig:fanopar}, the width of the resonance 
is directly related to the Kondo temperature ($T_K$). The Fano parameter $q$ contains 
information about the symmetry of the lineshape. We denote by $\epsilon_{K}$ the energy at 
which the resonance is centered. As introduced in Sec. \ref{sec:basic_theory}, this parameter 
is associated with the energy position of the Kondo resonance and therefore to the occupation 
of impurity level $n_{d}$, eq. (\ref{eq:1AIM-occup}). Finally, $A$ is the amplitude of the Fano 
profile and $g_{\rm off}$ the conductance offset of the curve out of the resonance. 

Each realization of the contact leads to a slightly different configuration. The statistical 
distributions of the different parameters of the Fano equation described above will reflect 
the subtle differences in the electronic structure of each contact. We present below a 
statistical analysis of each of these parameters for hundreds of contacts of Co, Fe and Ni. 
This novel statistical analysis (since in our previous work\cite{Calvo_Nature} 
we only analysed briefly the shape and mean value of the Kondo Temperature)
together with theoretical considerations brings new insight 
into the physics of Fano-Kondo resonances. We present now the distributions of these parameteres 
and compare the results between materials.

\subsection{Kondo Temperatures}

As described in Sec. \ref{sec:basic_theory} the zero-bias resonances observed in the conductance 
characteristics of Kondo systems are directly related to the resonances developed in the spectral 
density of the system. As explained in the theory section, the width of this resonance is determined 
by the energy scale of the Kondo screening, the so-called Kondo temperature $T_K$. Thus the width of 
the observed Fano lineshapes must be proportional to the Kondo temperature of the system. More 
precisely, we define the Kondo scale as the half width of the Fano lineshape: $\Gamma_{K}=k_{B}T_{K}$. 
The width of the measured Kondo resonance is strongly affected by a finite temperature of the system: 
in addition to the standard thermal broadening of any differential conductance feature, the Kondo 
resonance presents an intrinsic thermal broadening.\cite{Nagaoka_temp} This results in a considerable 
extra broadening of the resonance at temperatures on the order of magnitude of $T_{K}$. In our case, 
since the width of our resonances excesses in more than an order of magnitude the experimental 
temperature of 4.2K
and the bias voltages used are low enough \cite{kroger_heating}
, we can disregard thermal effects and consider that we can extract the Kondo 
temperature for each contact directly from the width of the Fano resonance.

\begin{figure}[ht]
  \epsfig{width=\linewidth,figure=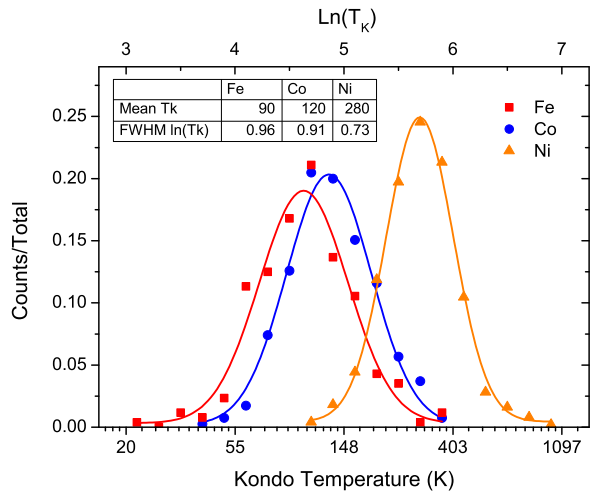}
  \caption{\label{fig:histotk}
    Histograms of values of logarithm of Kondo temperature for Fe, Co and Ni.
    The histograms are  normalized to the total number of counts. 
    The distribution of values for the ln(Tk) is clearly narrower for Ni than
    for Fe and Co in spite of the higher value of Kondo temperatures for Ni, 
    as summarized in the inset. The most frequent value of the Kondo temperature is determined with an error of 10 K.
  }
\end{figure}

As we have already described in Ref. \onlinecite{Calvo_Nature}, the distribution of Kondo temperatures 
fits a logarithmic normal distribution for the three materials Co, Fe and Ni, meaning that the logarithm
of $T_{K}$ is normally distributed 
(presented in Fig. \ref{fig:histotk}). This peculiar behavior is easily understood when interpreted in terms of the Kondo
effect: Since many different atomic configurations result in single-atom contacts, their electronic properties, such as 
conductance (Fig. \ref{fig:traces}), density of states and the associated energy scales are expected to be normally 
distributed. On the other hand following eq. (\ref{eq:tk}) the Kondo temperature depends \emph{exponentially} on the 
typical energy scales of the problem. Hence $\ln T_K$ for different contacts should follow a normal distribution if
the relevant energy scales of the problem are normally distributed.

By just looking at the resonances, as for example the ones shown in Fig. \ref{fig:spectroscopy},
it can be observed that the Fano features in the 
conductance spectra of Ni contacts are considerably broader than the ones for the case of Co and Fe. As shown in Fig. 
\ref{fig:histotk} and summarized in table \ref{tab:chemicaltrend},  the histograms yield most frequent values for the 
resonance widths of $T_{K}=90$~K, 120~K and 280~K for Fe, Co and Ni respectively, following the same trend 
$T_{K}^{Fe}<T_{K}^{Co}<T_{K}^{Ni}$ as in the case of adatoms of these elements deposited on non-magnetic surfaces 
\cite{Jamneala_FeCoNi} and diluted alloys of Cu containing the same concentration of magnetic atoms\cite{Daybell_review}. 
In simple terms, the Kondo temperature decreases as the size of the screened magnetic moment increases, as we go from Ni 
to Fe. 

\begin{table}[htp!]
  \centering
  \setlength{\tabcolsep}{3mm}
  \begin{tabular}[c]{l|c|c|c} 
    & \textbf{Fe} & \textbf{Co} & \textbf{Ni} \\ 
    \hline
    $T_{K}$ (K) in bulk Cu \cite{Daybell_review} & 10-50 &300-700 & $\simeq$1000 \\ 
    \hline
    $T_{K}$ (K) adatoms & --- & 53-92 \cite{Wahl_adatom}  & 120 \cite{Jamneala_FeCoNi}\\ 
    \hline
    $T_{K}$ (K) this work & 90 & 120 & 280   \\  
    \hline
    $m_{\rm atom}(\mu{B})$ & 3 & 2 & 1  \\  
    \hline
    $m_{\rm bulk}(\mu{B})$ & 2.22  & 1.72  & 0.60     \\  
  \end{tabular}
  \caption{\label{tab:chemicaltrend}
     A comparison of Kondo temperatures in the case of magnetic 
     impurities in bulk Cu, magnetic adatoms in tunnelling regime 
     and for the ferromagnetic contacts (Fe, Co and Ni) in this work.  
     The different values of Kondo temperatures in the case of 
     adatoms correspond to measurements performed over different substrates.
     Also shown are the magnetic moments of isolated atoms and in bulk.
  }
\end{table}

Figure \ref{fig:histotk} shows the distribution for $\ln(T_{K})$ fitted to a Gaussian distribution. Surprisingly 
enough, in spite of showing quite higher values of Kondo temperatures, the distribution of Ni when plotted in 
logarithmic scale is considerably narrower than in the case of Fe and Co (see also inset of Fig. \ref{fig:histotk}).
This suggests that in the case of Ni the characteristics of the Kondo screening are very different from the 
cases of Co and Fe possibly indicating a different mechanism for the case of Ni and the cases of Co and Fe.

The higher Kondo temperatures for Ni as well as their narrower distribution could well be connected to the
different chemical valence and the resulting magnetic moment in comparison to Co and Fe: While Ni basically 
has one hole in the $3d$-shell and therefore features an atomic spin of 1/2, Co and Fe have two and three holes
in their $3d$-shells associated with atomic spins of 1 and 3/2, respectively. Hence in the case of Co and
Fe the possibility exists that the full atomic spin $S>1/2$ is only partially screened while the spin-1/2
in the case of Ni is fully screened by the conduction electrons. The cases of Co and Fe would then resemble
the situation of an underscreened Kondo lattice where the remaining unscreened spin couples ferromagnetically
to the spins on neighbouring atoms as discussed in Ref. \onlinecite{Perkins_EPL}.
As explained in the theory section \ref{sec:basic_theory} such an underscreened Kondo effect is characterized 
by sharper resonances in comparison to the normal fully screened Kondo situation, and hence results in lower 
Kondo temperatures in the analysis. In this sense the Kondo temperatures may be underestimated in the case of
Co and Fe.
We discuss this in detail in the discussion section \ref{sec:discussion} together with the distribution of other 
parameters.

\subsection{Resonance energy ($\epsilon_{K}$) and $d$-level occupation}

\begin{figure}[ht]
  \epsfig{width=8.5cm,figure=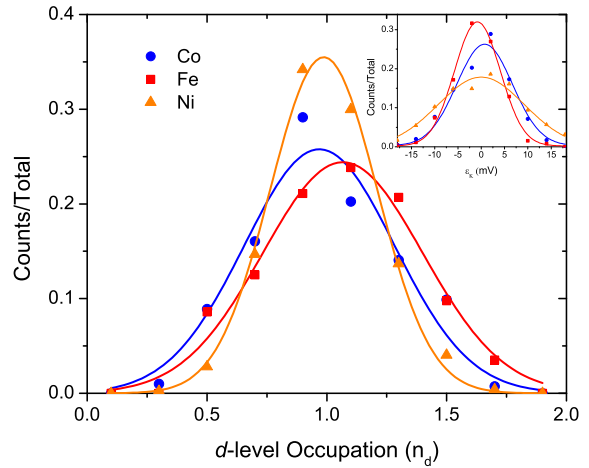}
  \caption{\label{fig:occupation} 
    Histograms of the $d$-level occupations $n_d$ for contacts of Fe, Co and Ni. The color lines show the fitting of these 
    distributions to a Gaussian peak. Surprisingly, the distribution for Ni is narrower than the other two, in spite of the 
    fact of this material showing a broader distribution of the parameter $\epsilon_{K}$.
    Inset: Distribution of $\epsilon_{K}$ for hundred of contacts of Fe, Co and Ni. This parameter accounts for the deviation 
    of the center of the Fano resonance from zero bias and is related to the position of the localized magnetic moment in the 
    Kondo model.
  }
\end{figure}

Another important parameter to study is the position of the Kondo resonance $\epsilon_{K}$ which accounts for the deviation 
of the center of the Fano resonance from zero bias and is related to the energy of the effective Kondo level with 
respect to the Fermi energy \cite{Wahl_adatom}. The inset of Fig. \ref{fig:occupation} shows the distribution of this 
parameter for hundred of contacts of the three materials under study. These distributions fit a Gaussian lineshape 
which is considerably broader for the case of Ni than for Co and Fe.

Most interestingly, from the values of $\epsilon_{K}$ and $T_{K}$ it is possible to extract the occupation of the 
$d$-level (giving rise to the Kondo resonance) from the experimental data, assuming a Fermi liquid approximation
as explained in Sec \ref{sec:basic_theory}, eq. (\ref{eq:1AIM-occup}). This approximation should be valid here 
since we are well below the Kondo temperature of our system. As explained in Sec. \ref{sec:basic_theory},
this relation is even valid for the underscreened Kondo effect in the multi-orbital Anderson model (although 
strictly speaking we then have a {\it Singular} Fermi liquid) where now $n_d$ refers 
to the the $d$-level whose spin is 
screened by the conduction electrons, and $\epsilon_K$ and $T_K$ refer to the position and width of the Kondo
resonance of the underscreened Kondo effect. Fig. \ref{fig:occupation} shows the distribution of positions and the 
resulting calculated occupations $n_d$ for Fe, Co and Ni which again follow a normal distribution. 

As summarized in Table \ref{tab:occupation} the mean value for the three materials is close to 1. Interestingly, 
in spite of the broader distribution of values of $\epsilon_{K}$, the occupations for Ni contacts clearly show  
a narrower distribution. Numerically this can be explained by the much narrower distribution of $ln(T_{K})$.
Physically, the reason behind the narrower distribution of Kondo temperatures and occupations in the case of
Ni might be that charge fluctuations become stronger with an increasing number of active levels as already
discussed in the seminal work of Nozieres:\cite{Nozieres_RealKondo}
In the case of Ni we are most likely dealing with a single active impurity level due to a single hole in the
$3d$-shell of Ni. Hence the charge in this level is quite well defined, and the occupation very close to one 
electron. In the cases of Co and Fe on the other hand we should have more than one active impurity level and therefore
the variation in the occupation of the level giving rise to the Kondo resonance is much stronger. Relatedly,
a broader distribution in the occupation of the impurity level giving rise to the Kondo resonance should also
give rise to a broader distribution in the Kondo temperatures since charge fluctuations strongly alter the width
of the Kondo resonance.\cite{Hewson}

\begin{table}[h]
  \begin{center}
    \setlength{\tabcolsep}{3mm}
     \begin{tabular}{l|c|c|c}
      & {\bf Fe} & {\bf Co} & {\bf Ni} \\
      \hline
      Mean $\epsilon_{K}$ (mV) & -1.0 & 0.6 & 0.3 \\
      \hline
      FWHM $\epsilon_{K}$(mV) & 14.7 & 16.3 & 25.4 \\
      \hline
      Mean $n_{d}$ & 1.07 & 0.97  &  0.99 \\
      \hline
      FWHM $n_{d}$ & 0.79  & 0.74  &  0.52  \\
    \end{tabular}
    \caption{\label{tab:occupation}
       Most frequent value of Resonance energy and d-level occupations and their respective 
       width of their Gaussian distributions for Fe, Co and Ni}
  \end{center}
\end{table}

\subsection{Amplitudes}

\begin{figure}[ht]
  \epsfig{width=\linewidth,figure=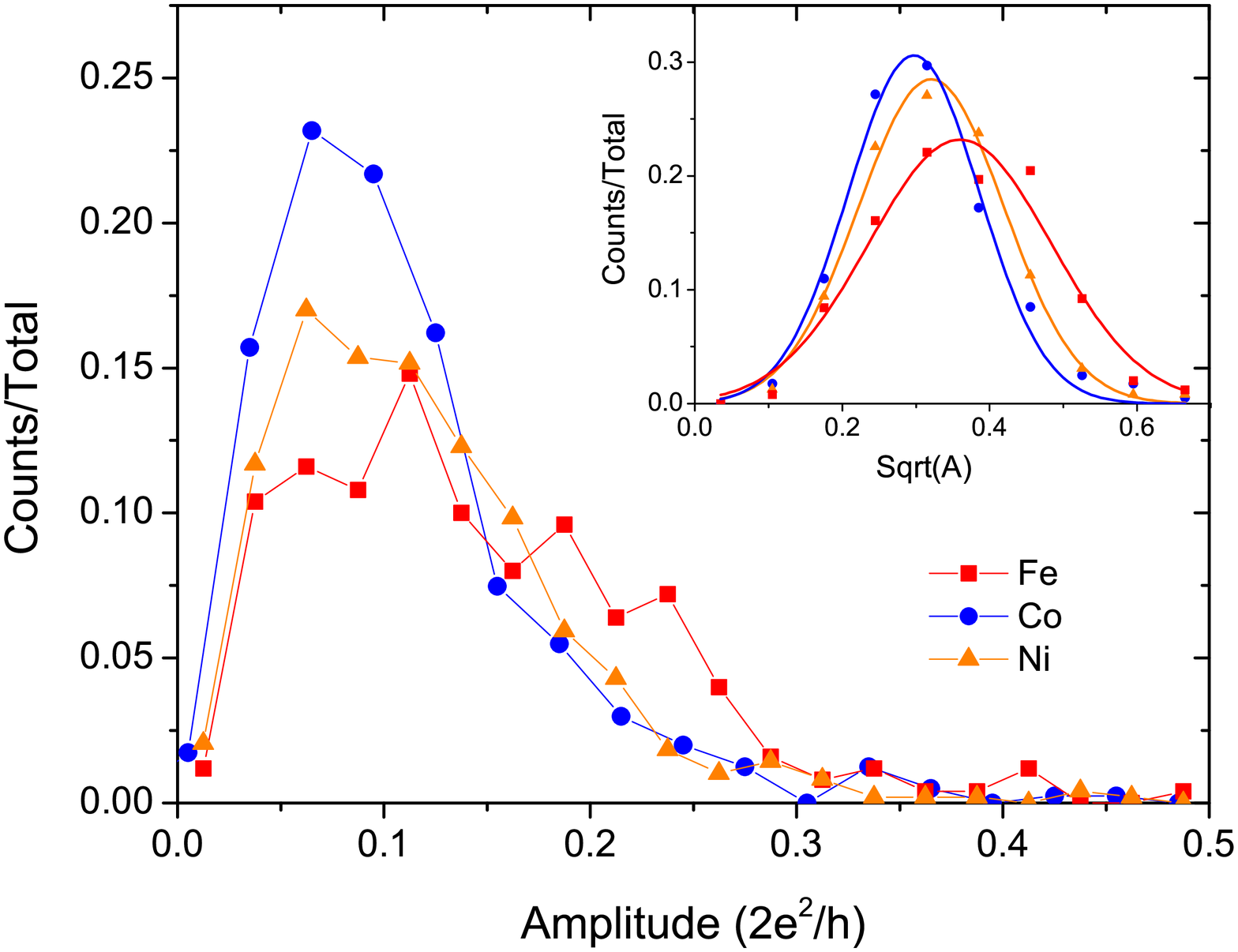}
  \caption{\label{fig:amp}
    Distribution of amplitudes extracted from the fitting of characteristics of hundred of 
    contacts of Fe, Co and Ni to the Fano equation. The distributions are similar for the 
    three materials, being the amplitudes about a 10 percent of the conductance of the contacts. 
    The inset shows the distribution of $\sqrt(A)$ and its fit to a Gaussian. This distribution 
    could reflect the quadratic dependence of amplitude in different coupling terms.}
\end{figure}

At first glance, the distribution of amplitudes in Fig. \ref{fig:amp} shows no clear differences 
between the three materials. The most frequent value of the amplitude is approximately 0.1~$2e^2/h$, 
i.e. about 10 percent of the conductance of the contact. 
On the other hand we find that the square root of the amplitude indeed follows a normal distribution
as can be seen from the inset of Fig. \ref{fig:amp} which shows the statistical distributions of the 
square root of the amplitude for the three materials. A possible explanation is that following expression
(\ref{eq:fano-A}) the amplitude depends quadratically on the coupling $V_{sd}$ between the $s$- and $d$-level
of the tip atom. $V_{sd}$ is expected to vary strongly when the atomic configuration of the contact changes
since it is induced by disorder in the contact region, and is absent for perfect crystalline order.
Hence if $V_{sd}$ is normally distributed and is the parameter determining $A$ that is most strongly affected 
by changing the atomic configuration of the contact one would expect $\sqrt{A}$ to be normally distributed.

Furthermore one can seen that the distributions of $\sqrt{A}$ for the three materials 
are centred at quite similar values and also have similar widths. However, we can make out a subtle 
trend (Tab. \ref{tab:amplitude}): The mean amplitude and the mean value of the square root of the amplitude 
both are slightly higher for Fe contacts than for the other two materials, and also both distributions
are slightly broader for Fe than for Co and Ni.
However, this trend is not as clear as the trend observed in the distributions of Kondo temperatures $T_{K}$
for the three materials (Fig. \ref{fig:histotk}). Thus it is difficult to draw any further conclusions from it. 
Moreover. we would like to point out that a similar trend is observed in the distributions of the conductances
(see Fig. \ref{fig:traces} and Tab. \ref{tab:amplitude}): Fe has a higher average conductance than Ni and Co.
More data would be needed to extract further conclusions from this analysis.


\begin{table}[h]
  \begin{center}
    \setlength{\tabcolsep}{3mm}
    \begin{tabular}{l|c|c|c}
      & {\bf Fe} & {\bf Co} & {\bf Ni} \\
      \hline
      Mean $A$ from $\sqrt{A}$ & 0.13 & 0.088  &  0.10  \\
      \hline
      Mean $\sqrt{A}$ & 0.36  & 0.30 & 0.32   \\
      \hline
      FWHM $\sqrt{A}$ & 0.31   & 0.24   & 0.27   \\
      \hline
      Mean $G$ from Fig. \ref{fig:traces} & 2.0 & 1.2  & 1.6 \\ 
    \end{tabular}
    \caption{\label{tab:amplitude}
      Mean values of amplitudes and width of the
      the gaussian distributions of $\sqrt{A}$ and 
      mean conductances for Fe, Co and Ni.}
  \end{center}
\end{table}

\begin{figure}[htp]
  \centering
  \includegraphics[width=\linewidth]{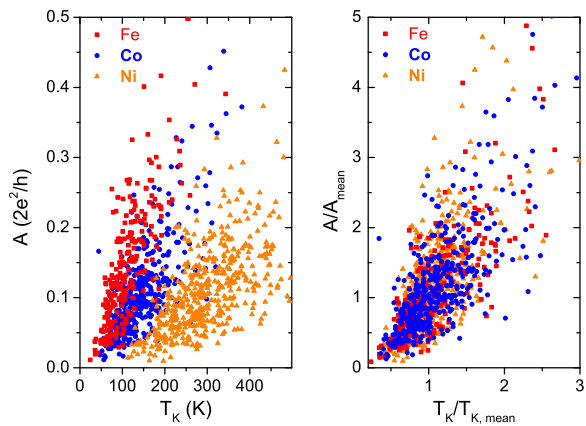}
  \caption{\label{fig:ampvstk}
    Left: Scattered plot of the amplitude of the resonance versus the Kondo temperature obtained 
    for hundreds of contacts of Fe, Co and Ni. The plot shows a clearly similar (linear) trend 
    for the three materials. When divided by the most probable value of the Kondo temperature 
    for each material, the distributions lay over each other, as shown in the right panel.
  }
\end{figure}

Far more interestingly, the plot of the amplitude versus the Kondo temperature presents an intriguing trend
as can be seen from Fig. \ref{fig:ampvstk}: For all three materials there seems to be an approximately
linear dependence of the amplitude on the Kondo temperature with respect to the Kondo temperature.
Moreover, when normalizing the amplitudes
and Kondo temperatures to the respective mean values (Tab.\ref{tab:amplitude}) one obtains a {\it universal}
dependence suggesting that $A\propto f(T_K)$.

This {\it universal} behaviour 
seems to reflect some kind of universality in the relation between the basic parameters $U$, $\epsilon_d$ 
and $\Gamma_d$ that ultimately determine the Kondo properties of our system.
Said in another way, the basic parameters $U$, $\epsilon_d$ and $\Gamma_d$ are not independent from each 
other but are linked together in such a way that universal scaling between the amplitude $A_K$ and the
Kondo temperature $T_K$ results.
For example it is conceivable that both $\Gamma_d$ (the coupling of the impurity level to the conduction
electrons) and $U$ (the effective Coulomb repulsion of the impurity level) are related since a change
in the coupling $\Gamma_d$ implies a change in the localization of electrons in the impurity level and
hence can result in an alteration of the screening of the effective Coulomb interaction $U$.
Further theoretical work is necessary in order to achieve a rigorous interpretation of these results.

\subsection{Fano Parameters}
\label{sec:fano}

\begin{figure}[htp]
  \centering
  \includegraphics[width=\linewidth]{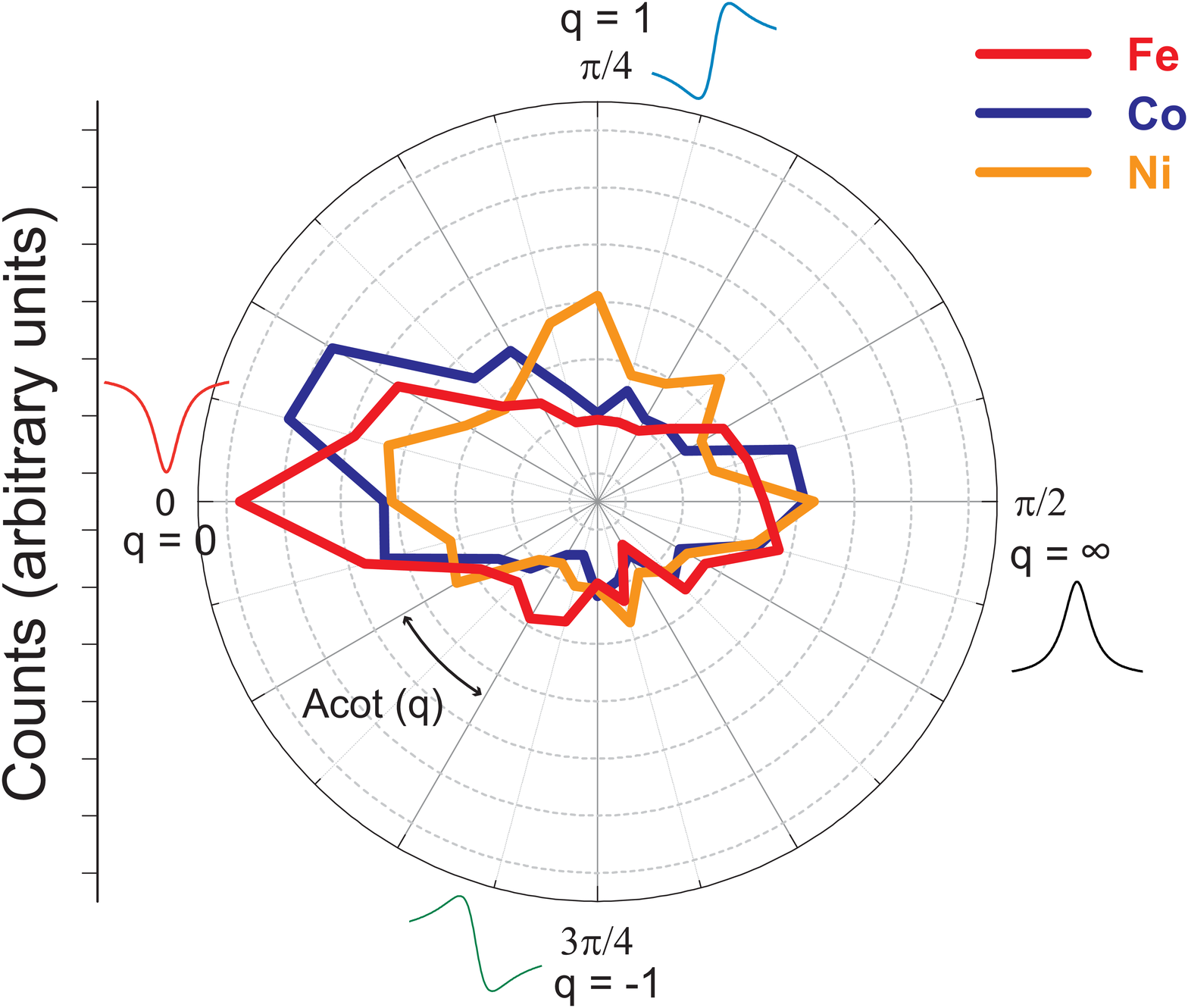}
  \caption{\label{fig:fanos}
    The so-called Fano parameter $q$ ranges from $0$ to $\infty$, but it can be expressed as 
    $q=\tan(\alpha)$ with $\alpha$ between 0 and $\pi$, for simplicity in the representation. 
    $\alpha=0$ corresponds to the deep, $\alpha=\pi$ to the peak and $\alpha=1,-1$ the totally 
    asymmetric cases. The histograms are normalized to the total number of data.
  }
\end{figure}

The Fano parameter $q$ accounts for the symmetry of the Fano resonances, recovering the perfect 
Lorentzian shape for $q\rightarrow\infty$ and its inverse for $q=0$. 
As already commented above all possible symmetries 
are found for each material. A histogram of the values of $q$ for the three materials is shown in 
Fig. \ref{fig:fanos}. Since $q$ ranges from 0 to $\infty$, an alternative representation where 
$q=\tan(\alpha)$ is chosen for simplicity. In this representation $\alpha=0$ corresponds to the 
dip and $\alpha=\pi$ to the peak lineshapes. The totally asymmetric cases are represented by 
$\alpha=\pi/2$ ($q=1$) and $\alpha=3\pi/2$ ($q=-1$).
There is a remarkable preference towards the symmetric cases over the more asymmetric ones in 
the case of Fe and Co, and in particular the dip-like ($q=0$) lineshapes seem to prevail. 
For Ni, the occurrence of asymmetric cases ($|q|\sim1$) is considerably higher, overall the 
Fano factors seem to be more evenly distributed than in the case of Fe and Co. 

In the case of adatoms in tunnelling, the symmetry is well understood in terms of the ratio between the 
probability of transmission through the $d$ or the $s$ channels\cite{Ternes_review}. Nevertheless, in 
the point contact regime and in our case, the system presents a higher complexity. There is a totally 
open $s$-channel and the nature of the orbitals and couplings contributing to the interference is not 
completely clear. The distribution of the $q$ parameters should contain information about the contributions 
from the different channels to the interference which might be extracted with the help of an appropriate 
theoretical treatment.

\begin{figure}
  \includegraphics[width=\linewidth]{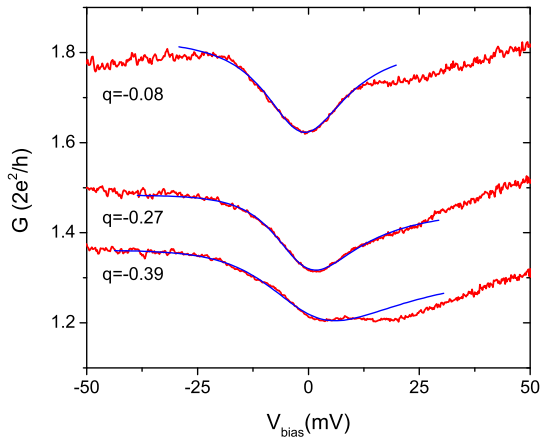}
  \caption{\label{fig_stress}
    Evolution of the dI/dV characteristics of a single Co contact as stretching the junction. 
    As the conductance of the contact decreases, also the symmetry of the curve, that is the 
    Fano parameter, $q$ changes as a consequence of the different electronic structure of 
    different arrangements.
  }
\end{figure}

In Fig. \ref{fig_stress} we plot the evolution of a single Co contact when stretching the junction. The 
symmetry of the curve changes as the contact rearranges in a different manner which translates into 
a different conductance of the curve (the curves are not offset). When stretching the contact the 
the hoppings between different orbitals, the electronic structure and hence the conductance varies,
this affects the interference between the conduction channels and also the Kondo resonance responsible 
for the Fano lineshapes. This evolution is probably specific for each contact since an attempt to find 
a statistical relation between $q$ and the conductance offset results in a random distribution of points. 
More work is needed to fully clarify this effect.

\subsection{Summary}

In summary, the statistical analysis of the Fano-Kondo lineshapes extracted from hundreds of 
atomic contacts shows that:

(i) The statistical distribution of Kondo temperatures for each material is a log-normal 
distribution, i. e. the logarithm of the Kondo temperatures is normally distributed. This 
is expected in the context of the Kondo effect since the Kondo temperature depends 
exponentially on the different electronic properties and couplings of the system. These 
properties are likely to be normally distributed for the slightly different atomic 
configurations of the contacts. Hence the log-normal distribution of Kondo temperature 
presents further evidence that the Kondo effect is responsible for the zero-bias anomalies 
in the conductance.

(ii) The mean value for the Kondo temperatures is considerably higher for Ni than for Co and 
Fe whereas the distribution of the logarithm of the Kondo temperature for Ni is narrower than 
for Co or Fe. This finding suggests that possibly a different mechanism is responsible for 
the Kondo screening in the case of Ni than in the case of Co and Fe.

(iii) The distribution of the resonance energies $\epsilon_{K}$ is Gaussian and is centred 
around zero for all three materials.  In the case of Ni this distribution is much broader 
than in the case of Fe and Co where the distribution widths are similar. This again points 
to different mechanisms for the Kondo screening for Ni on the one hand and for Co and Fe on 
the other hand.

(iv) The occupations $n_d$ of the impurity level giving rise to the Kondo resonance calculated 
from $T_K$ and $\epsilon_K$ by eq. (\ref{eq:1AIM-occup}) again follow a Gaussian distribution 
centered around occupation 1. This distribution is narrower for Ni than for Co and Fe. This can 
be understood by considering that most likely for Ni only a single $d$-level is active while for 
Fe and Co several $d$-levels must be active. This leads to stronger charge fluctuations and 
hence a larger variation in the $d$-level occupations in the case of Co and Fe.

(v) The distribution of the Fano curve amplitudes does not show significant differences between 
the materials. We find that the square root of the amplitudes is normally distributed. This again
can be understood by a normal distribution of the characteristic parameters of the nanocontacts 
on which the amplitude depends quadratically.

(vi) The plot of the amplitudes against the Kondo temperatures follows a similar (almost linear) 
trend for the three materials. When divided by the mean values for each material the data lines 
are perfectly on top of each other, showing a universal scaling behaviour of the amplitude with 
the Kondo temperature. 

(vii) There are more asymmetric Fano lineshapes in the case of Ni than for Co and Fe. The 
distribution of Fano parameters $q$ again is similar for Fe and Co but different for Ni. Co
 and Fe show a clear preference for the dip-like $q=0$ line shapes while for Ni the distribution 
of the Fano parameters is somewhat more uniform.

Table \ref{tab:summary} shows a summary of the parameters extracted from the statistical 
analysis of the Fano-Kondo lineshapes for the all three materials.

\begin{table}[h]
  \begin{center}
    \setlength{\tabcolsep}{3mm}
    \begin{tabular}{l|c|c|c}
      & {\bf Fe} & {\bf Co} & {\bf Ni} \\
      \hline
      Mean $T_{K}$ & 90 & 120  & 280  \\
      \hline
      FWHM $ln(T_{K})$& 0.96 & 0.91  & 0.73   \\
      \hline
      Mean $\epsilon_{K}$ (mV) & -1.0 & 0.6 & 0.3 \\
      \hline
      FWHM $\epsilon_{K}$(mV) & 14.7 & 16.3 & 25.4 \\
      \hline
      Mean $n_{d}$ & 1.07 & 0.97  &  0.99 \\
      \hline
      FWHM $n_{d}$ & 0.79  & 0.74  &  0.52  \\
      \hline
      Mean $A$ from $\sqrt{A}$ & 0.13 & 0.088  &  0.1  \\
      \hline
      FWHM $\sqrt(A)$ & 0.31   & 0.24   & 0.27   \\
    \end{tabular}
    \caption{\label{tab:summary}
      Summary of the different parameters extracted from 
      the statistical analysis of hundreds of atomic contacts
      for each of the three materials Fe, Co and Ni.
    }
  \end{center}
  \vspace{-0.6cm}
\end{table}

\section{Discussion}
\label{sec:discussion}

The picture of the Kondo effect in ferromagnetic atomic contacts that emerges from our 
statistical analysis and theoretical considerations is the following: The low coordination 
and disorder of the atoms in the contact region in connection with a higher effective Coulomb 
repulsion can lead to the localization of a single spin in an individual $d$-level of an atom 
in the contact region.\cite{Calvo_Nature}
Now, due to disorder in the contact region it can happen that this $d$-level only couples
very weakly to the spin-polarized $d$-levels on neighbouring atoms but instead it has an
effective coupling to the basically spin-unpolarized $s$-type conduction channel. 
In this situation the Kondo effect can take place and screen the spin in that $d$-level.

In the case of Ni there is one hole in the $3d$-shell and therefore a spin-1/2 associated 
with it. If the spin becomes localized in a $d$-level that predominantly couples to the 
spin-unpolarized $s$-type conduction electrons this spin can be fully screened. Hence for 
Ni we should have a {\it normal} Kondo effect where the full spin-1/2 is screened. In the 
case of Fe and Co the atomic spin is higher due to 3 and 2 holes, respectively, in the 
$3d$-shell of these atoms. This might explain why the measured Kondo temperatures are higher 
in the case of Ni than for Co and Fe since the Kondo temperature decreases with increasing 
spin of the impurity.\cite{Goldberg,Nevidomskyy} 
However, one would then also expect a significantly lower Kondo temperature for Fe than for 
Co which is not the case.

The scenario that we propose instead for Co and Fe is an underscreened Kondo effect where the 
full atomic spin $S>1/2$ is only partially screened by the conduction electrons. More precisely, 
only the spin-1/2 within the $d$-level that predominantly couples to the $s$-type conduction 
electrons will be screened while the rest of the spin $S-1/2$ which is likely to be localized 
in $d$-levels that couple more strongly to the spin-polarized $d$-levels on neighbouring atoms, 
remains unscreened. This scenario fits very well with our results: In particular, it explains 
why the average Kondo temperature is very similar for Fe and Co but is considerably higher for 
Ni as the underscreened Kondo effect is characterized by a cusp-like resonance much sharper than
the Lorentzian-type resonance of the normal fully screened Kondo effect. An underscreened Kondo
effect for Fe and Co contacts would also explain why the occupations can be calculated from eq.
(\ref{eq:1AIM-occup}) derived for the single-level Anderson impurity model. 


A recent work by N\'eel {\it et a}.\cite{Berndt_Fetip} shows how the Kondo temperature increases for a Cobalt impurity when it is contacted by an Fe tip compared to the case of using a Cu tip. These results and their interpretation are in good agreement with the results presented here: differences in the composition and geometries of the junctions lead to different electronic structure of the contact which determine the Kondo screening. By studying a specific system with a well characterized geometry, Neel et al. can extract the change in hybridization and the corresponding occupation of the d-level as the main cause of the observed increase in Kondo temperatures. We study here instead a large variety of possibilities. Changes in the hybridization are probably responsible for the broad distribution of Kondo temperatures in our data, but other parameters such as the occupancy of the $3d$-shell may also change in our system from contact to contact.


Finally, we would like to discuss the connection to a related work by Bork {\it et 
al.}\cite{bork_twoimpurities} with the results presented here: In the work of Bork 
{\it et al.} the Fano-Kondo lineshapes of a system consisting of two Co atoms, one on 
a Cu surface and the other attached to Cu STM tip are recorded while the distance between 
the two Co atoms is decreased from the tunnelling regime to the contact regime. The authors 
report a splitting of the Fano-Kondo lineshapes when the contact regime is entered due to 
the formation of a spin-singlet state between the two Co atoms. 

In some cases the Fano-Kondo lineshapes of the ferromagnetic contacts measured here could
also be interpreted as a splitting of the Kondo resonance. For example, the evolution 
of the Fano lineshape of a Co nanocontact being stretched (Fig. \ref{fig_stress}) possibly
shows a small splitting for the curve with the lowest conductance. However, if this specific 
feature really shows a splitting of the Fano-Kondo resonance it actually occurs in the opposite 
direction as for the system studied by Bork {\it et al.}, i.e. it occurs when the nanocontact
is pulled apart and not when the tip atoms are brought together as in the case of Bork {\it et al.}
In any case we have not observed such curves that could be interpreted as splittings 
very frequently, and the focus of this work
is a statistical analysis of a large number of different configurations of atomic contacts while 
the work of Bork {\it et al.} focuses on a very specific system. 
On the other hand we also would like to point out 
that conductance oscillations in the spectroscopy at the atomic scale\cite{Untiedt_PRB00} 
further complicate the interpretation of this kind of curves.

Still one might wonder why splitting of the Fano-Kondo lineshapes in the atomic contacts studied
here are not observed more frequently. The reason might be that indeed the 
$d$-level on a tip atom giving rise to the Kondo effect just does not couple very well to the 
$d$-levels of the other tip atom due to the disorder in the contact region. Hence the splitting
is probably very weak compared to the Kondo temperature and therefore not observed,
as also shown in Ref. \cite{Berndt_Fetip}, where the Kondo resonance remains unsplitted 
even when the Co adatom is contacted by a ferromagnetic tip.
Another possible explanation could be that the weak coupling between the $d$-levels giving rise
to the Kondo effect localized on different tip atoms is compensated by the weak spin polarization
of the conduction electrons similar to the case of a double quantum dot coupled to ferromagnetic 
leads.\cite{Zitko}

\section{Conclusions}
\label{sec:conclusions}

Atomic contacts are a unique system to understand the dramatic consequences of low coordination 
and disorder on the electronic transport and magnetic properties at the nanoscale. The emergence 
of the Kondo effect in atomic contacts is a good example for this.\cite{Calvo_Nature} 
Here we have extended our previous work in several aspects in order to gain new insights
into the nature of the Kondo effect in ferromagnetic one-atom contacts:
We have presented an exhaustive statistical analysis of the Fano resonances in 
the spectroscopy of ferromagnetic one-atom contacts. In particular, we have analysed 
not only the distribution of Kondo temperatures, but also of the resonance energies, 
amplitudes and Fano parameters. Such an exhaustive statistical analysis has never been 
performed before for any Kondo system, and it provides us with new information on the nature 
of the Kondo effect in these system. 

From this analysis we have for example obtained the distribution of $d$-level 
occupations, and the dependence of the amplitudes of the Fano resonance on the Kondo
temperature. We also point out the widths of the parameter distributions 
as an indicator on the robustness of the Kondo screening in one-atom contacts of the 
different materials.  Additionally, we have further developed the theory in order to 
explain the observed qualitative differences between Ni on the one hand and Co and Fe 
on the other hand. Furthermore, we have devised a simple microscopic model which allows 
us to understand the occurrence of different Fano lineshapes for the same material in 
terms of variations of the microscopic interactions.

Our statistical analysis shows clear differences in the Kondo characteristics of different 
ferromagnetic materials, i.e. of Fe and Co versus Ni. These differences can be explained by 
the different valences, leading to different Kondo scenarios. 
Although this might not be a unique explanation for the observed phenomenology, the reported 
results seem to fit well with an underscreened Kondo effect for the case of Fe and Co, and a 
standard spin-1/2 Kondo model for the case of Nickel.

While the different screening conditions might account for the differences in Kondo temperatures 
and occupation, the complexity of atomic contacts, especially of transitions metals, where many 
different channels contribute to the transport, makes it difficult to fully understand the 
details for each single contact. A simple interpretation for certain parameters as the exact 
shape of the curves (Fano parameter) or the amplitude of the resonances is difficult to achieve 
as the spectroscopy might contain also information about other phenomena such as quantum 
oscillations due to interferences between non resonant conduction channels, inelastic processes 
as phonon excitations etc. 

In the recent past, the combination of molecular dynamics simulations and {\it ab initio} 
transport calculations and experiment has proven to be a successful path for understanding 
the atomic, electronic and magnetic stucture as well as the transport properties of 
nanocontacts.\cite{Calvo_ieee,Untiedt_PRL07,annealing}
Further work in this direction in combination with more sophisticated many-body techniques 
capable of describing the Kondo effect such as the Dynamical Mean-Field 
Theory\cite{Jacob_Kondo,Jacob_DMFT} should contribute to fully understand Kondo physics and 
the presence of magnetism in these systems and in atomic-size structures in general.

\section{Acknowledgements}

This work was partly supported by the Spanish MEC (grants FIS2010-21883-C02-01 and CONSOLIDER 
CSD2007-0010) and Comunidad Valenciana (ACOMP/2012/127 and PROMETEO/2012/011). 
M. R. Calvo aknowledges finantial support from the Spanish MEC through the postdoctoral fellowships program. We are grateful to Vicent Esteve for 
technical support and J.I. Fern\'andez-Rossier, J.J. Palacios and 
D. Natelson for stimulating scientific discussions.

\appendix

\section{Two-level Anderson impurity eigenstates}
\label{app:2AIM}

Here we briefly discuss the many-body eigenstates of the {\it isolated} two-level
Anderson impurity model and the fluctuations between them that lead to the underscreened
Kondo effect as discussed in Sec. \ref{sec:UKE}.

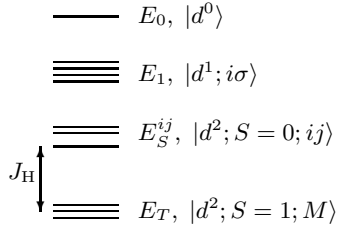
\begin{figure}
  \setlength{\unitlength}{\linewidth}
  \begin{picture}(0.5,0.4)
    \put(0.09,0.35){\line(1,0){0.1}} \put(0.22,0.34){$E_0,\;|d^0\rangle$}
    \put(0.09,0.28){\line(1,0){0.1}}
    \put(0.09,0.27){\line(1,0){0.1}}
    \put(0.09,0.26){\line(1,0){0.1}} \put(0.22,0.25){$E_1,\;|d^1;i\sigma\rangle$}
    \put(0.09,0.25){\line(1,0){0.1}}
    \put(0.09,0.18){\line(1,0){0.1}}
    \put(0.09,0.17){\line(1,0){0.1}}
    \put(0.09,0.15){\line(1,0){0.1}} \put(0.22,0.16){$E_S^{ij},\;|d^2;S=0;ij\rangle$}
    \put(0.09,0.06){\line(1,0){0.1}}
    \put(0.09,0.05){\line(1,0){0.1}} \put(0.22,0.04){$E_T,\;|d^2;S=1;M\rangle$}
    \put(0.09,0.04){\line(1,0){0.1}}
    \put(0.07,0.05){\vector(0,1){0.1}}
    \put(0.07,0.15){\vector(0,-1){0.1}}
    \put(0.02,0.10){$J_{\rm H}$}
  \end{picture}
  \caption{
    \label{fig:2AIM-levels}
    Schematic energy level diagram for the (isolated) 2-level impurity
    described by $H_{\rm I}$ in eq. (\ref{eq:2AIM}). }
\end{figure}

The isolated impurity Hamiltonian $\hat{H}_{\rm I}$ of the 2AIM (\ref{eq:2AIM}) 
is easily diagonalized. The empty impurity state is denoted by $\ket{d^0}$ and has 
energy $E_0=0$. For single occupation the four eigenstates and energies are 
trivially given by
\begin{equation}
  \ket{d^1;i\sigma} = d_{i\sigma}^\dagger\ket{0}, E_1 =\bra{i\sigma}\hat{H}_I\ket{i\sigma} = \epsilon
\end{equation}
For double occupation, the eigenstates can be separated into spin-singlet 
(total impurity spin $S=0$) and spin-triplet ($S=1$) states. 
The latter are given by 
\begin{eqnarray}
  \ket{d^2;S=1;M} = \ket{1,2}^- \otimes 
  \left\{
    \begin{array}{ll} 
      \ket{\su,\su}   & (M=1) \\
      \ket{\su,\sd}^+ & (M=0) \\
      \ket{\sd,\sd}   & (M=-1) 
    \end{array}
    \right.
\end{eqnarray}
where $M$ is the projection of the total impurity spin
onto the spin quantization axis.
The triplet states are all degenerate with energy
$E_T=2\epsilon+U^\prime-J_{\rm H}/4$.
The spin singlet states can be written as:
\begin{eqnarray}
  \ket{d^2;S=0;ij} &=& \ket{i,j}^+ \otimes \ket{\su,\sd}^-
\end{eqnarray}
The corresponding eigen energies are $E_S^{12}=2\epsilon+U^\prime+\frac{3}{4}J_{\rm H}$
and $E_S^{11}=E_S^{22}=2\epsilon+U$. Typically, $U\approx U^\prime+J_H$,
and therefore $E_S^{11}=E_S^{22}>E_S^{12}$. 
Due to Hund's rule coupling the triplet states are lower in energy by an
amount $E_S^{12}-E)_T=J_{\rm H}$. In Fig. \ref{fig:2AIM-levels} we show
the corresponding energy level diagram for the 2AIM. $U$ and $U^\prime$ are assumed to
be big enough to prevent triple and full occupation of the impurity.
Hence the triplet states comprise the ground state manifold of the isolated impurity
system.

Now in the situation of the underscreened Kondo effect where only one of the impurity levels 
is coupled to the conduction electron bath, hopping processes between the impurity and the 
bath can only lead to a spin flip in the impurity level that is coupled to the bath as 
illustrated in Fig. \ref{fig:UKE}. These spin flip processes can only lead to fluctuations
between states with $M=1$ and $M=0$ and between states with $M=-1$ and $M=0$. For example:
$\ket{S=1;M=1}=\ket{1\uparrow,2\uparrow}^{(-)}\stackrel{d_{1\uparrow}}{\longrightarrow}\ket{2\uparrow} 
\stackrel{d_{1\downarrow}^\dagger}{\longrightarrow}\ket{1\downarrow,2\uparrow}^{(-)}
\stackrel{d_{1\downarrow}}{\longrightarrow}\ket{2\uparrow}
\stackrel{d_{1\uparrow}^\dagger}\longrightarrow\ket{1\uparrow,2\uparrow}^{(-)}$.
Hence the total spin of the impurity is only {\it partially} screened by these processes.
This is the essence of the underscreened Kondo effect.

\section{Derivation of the Fano formula for model contact}

We assume the simplified model of a nanocontact described in Sec. \ref{sec:basic_theory}
and shown in Fig. \ref{fig:model}. We concentrate on one of the tip atoms of the nanocontact
which in our model consists of an (almost) perfectly transmitting $s$-level and the $d$-level
where the Kondo effect takes place. Hence the Greens function (GF) of the tip atom can be 
written as:
\begin{equation}
  \label{eq:GFA}
  \mathbf{G}_A(\omega) = \left(
  \begin{array}{cc}
    G_s & G_{sd} \\
    G_{sd} & G_d
  \end{array}
  \right)
\end{equation}
$G_d$ is the Green's function of the $d$-level which in the Kondo regime is given by 
eq. (\ref{eq:Gd}) and thus yields a Kondo resonance in the corresponding spectral 
function as described by eq. (\ref{eq:rhod}) at energy $\epsilon_K$, half width $\Gamma_K$
and the quasi particle weight $z$.

The unperturbed $s$-level GF (i.e. without coupling $V_{sd}$ to the $d$-level) is given by
\begin{equation}
  \label{eq:Gs0}
  G_s^0(\omega) = \frac{1}{\omega-\epsilon_s+i\Gamma_s/2} \approx \frac{1}{-\epsilon_s+i\Gamma_s/2}
\end{equation}
where $\epsilon_s$ is the energy of the $s$-level and $\Gamma_s=\Gamma_{L,s}+\Gamma_{R,s}$ 
the width due to coupling to both electrodes. Since the energies we are interested in are 
on the order of the Kondo scale, $|\omega|\approx\Gamma_K$ and $\Gamma_K\ll\Gamma_s$, we 
neglect the $\omega$-dependence of $G_s^0$ in the last step of eq. (\ref{eq:Gs0}).

Now due to the coupling $V_{sd}$ of the $s$-level to the $d$-level, the full GF $G_s$ 
of the $s$-level is modified according to:
\begin{equation}
  \label{eq:Gs}
  G_s(\omega) = G_s^0(\omega) + G_s^0(\omega)\,V_{sd}\,G_d(\omega)\,V_{sd}\,G_s^0(\omega)
\end{equation}

The off-diagonal term $G_{sd}$ describes the interference between both channels due to the coupling
$V_{sd}$, and is given by:
\begin{equation}
  \label{eq:Gsd}
  G_{sd}(\omega) = V_{sd} \cdot G_s^0 \cdot G_d(\omega)
\end{equation}

As was shown by Meir and Wingreen in their seminal work\cite{Meir-Wingreen},
at zero temperature and in linear response the exact result for the
conductance and current through an interacting region reduces to the Landauer
result where the transmission function can be calculated from the Caroli
formula\cite{Caroli}. Using the Caroli formula we can calculate the 
coherent transmission via the tip atom as
\begin{equation}
  \label{eq:caroli}
  T(\omega) = {\rm Tr}[ \mathbf\Gamma_L \mathbf{G}_{A}^\dagger \mathbf\Gamma_R \mathbf{G}_{A} ]
\end{equation}
where $\mathbf\Gamma_L$ and $\mathbf\Gamma_R$ are the so-called coupling matrices describing the
coupling of the tip atom $A$ to the left and right electrodes:
\begin{equation}
  \mathbf\Gamma_\alpha = \left(
  \begin{array}{cc}
    \Gamma_{\alpha,s} & 0 \\
    0 & \Gamma_{\alpha,d}
  \end{array}
  \right)
  \hspace{1ex}\mbox{with}\,\alpha\in\{L,R\}
\end{equation}
Hence we find for the transmission $T(\omega)$ through the tip atom:
\begin{eqnarray} 
  \label{eq:fulltransm}
  T(\omega) = T_s(\omega) + T_d(\omega) + T_{sd}(\omega)
\end{eqnarray}
where $T_s$ is the direct transmission through the $s$-channel and $T_d$ the corresponding one
through the $d$-channel while $T_{sd}$ describes the transmission involving hopping  processes 
between the $s$- and $d$-channel:
\begin{eqnarray} 
  \label{eq:Ts}
  T_s(\omega) &=& 2\pi\cdot\frac{\Gamma_{L,s}\cdot\Gamma_{R,s}}{\Gamma_s}\cdot\rho_s(\omega)\\
  \label{eq:Td}
  T_d(\omega) &=& 2\pi\cdot\frac{\Gamma_{L,d}\cdot\Gamma_{R,d}}{\Gamma_d}\cdot\rho_d(\omega) \\
  \label{eq:Tsd}
  T_{sd}(\omega) &=& (\Gamma_{L,s}\Gamma_{R,d}+\Gamma_{L,d}\Gamma_{R,s})\cdot|G_{sd}|^2
\end{eqnarray}

In order to calculate the contribution of the $s$-channel to the total transmission, 
we need to know the spectral density of the $s$-level which is given by the imaginary part 
of the $s$-level GF given by eq. (\ref{eq:Gs}):
\begin{eqnarray}
  \rho_s(\omega) &=& \underbrace{-\frac{1}{\pi}{\rm Im}[G_s^0(\omega)]}_{\rho_s^0(\omega)}
  \underbrace{-\frac{V_{sd}^2}{\pi}\cdot{\rm Im}[(G_s^0(\omega))^2\cdot G_d(\omega)]}_{\delta\rho_s(\omega)}
  \nonumber\\
\end{eqnarray}
where $\rho_s^0$ is the spectral function of the unperturbed $s$-level which in our model is 
constant, $\rho_s^0=\Gamma_s/2\pi(\Gamma_s^2/4+\epsilon_s^2)$. $\delta\rho_s(\omega)$ is the 
change in the spectral density due to the coupling to the $d$-level with the Kondo peak.
For the latter we find:
\begin{eqnarray}
  \label{eq:deltarhos}
  \delta\rho_s
  &=& -\frac{V_{sd}^2}{\pi} \cdot \left\{ 
    {\rm Im}[(G_s^0)^2]\cdot{\rm Re}[G_d] + {\rm Re}[(G_s^0)^2]\cdot{\rm Im}[G_d]
  \right\} \nonumber\\
  &=& -\frac{V_{sd}^2}{\pi} \cdot \left\{ 
    2\cdot{\rm Re}[G_s^0]\cdot{\rm Im}[G_s^0]\cdot{\rm Re}[G_d] 
  \right.\nonumber\\
  && \left. 
    + ({\rm Re}[G_s^0]^2-{\rm Im}[G_s^0]^2) \cdot{\rm Im}[G_d] 
  \right\}
\end{eqnarray}
The real and imaginary part of $G_d$ are given by:
\begin{eqnarray}
  {\rm Re}\,G_d(\omega) &=& \frac{z\,(\omega-\epsilon_K)}{(\omega-\epsilon_K)^2+\Gamma_K^2} = \frac{z}{\Gamma_K}\cdot\frac{x}{x^2+1} \hspace{5ex} \\
  -{\rm Im}\,G_d(\omega) &=& \frac{z\,\Gamma_K}{(\omega-\epsilon_K)^2+\Gamma_K^2} = \frac{z}{\Gamma_K}\cdot\frac{1}{x^2+1} \hspace{5ex}
\end{eqnarray}
where we have defined $x\equiv(\omega-\epsilon_K)/\Gamma_K$.
Plugging this into eq. (\ref{eq:deltarhos}) we find:
\begin{eqnarray}
  \label{eq:deltarhos2}
  \lefteqn{\delta\rho_s = -\frac{z\,V_{sd}^2}{\pi\,\Gamma_K}\cdot\frac{1}{x^2+1} \cdot} \nonumber\\
  && \cdot\left\{ 2x\cdot{\rm Re}\,G_s^0\cdot{\rm Im}\,G_s^0 -[{\rm Re}\,G_s^0]^2 + [{\rm Im}\,G_s^0]^2\right\} 
\end{eqnarray}

We now define the ratio $q$ between real and imaginary part of $G_s^0$:
\begin{equation}
  q\equiv-\frac{{\rm Re}\,G_s^0}{{\rm Im}\,G_s^0} \approx -\frac{2\epsilon_s}{\Gamma_s}
\end{equation}
With this we find for the expression in curly brackets in eq. (\ref{eq:deltarhos2}):
\begin{eqnarray}
  \lefteqn{2x\cdot{\rm Re}\,G_s^0\cdot{\rm Im}\,G_s^0  -[{\rm Re}\,G_s^0]^2 + [{\rm Im}\,G_s^0]^2 =}\\
  &=& [{\rm Im}\,G_s^0]^2 \times \left\{ 
  2x\cdot \frac{{\rm Re}\,G_s^0}{{\rm Im}\,G_s^0} 
  -\left[\frac{{\rm Re}\,G_s^0}{{\rm Im}\,G_s^0}\right]^2 + 1 
  \right\} \nonumber\\
  &\approx& \frac{4}{\Gamma_s^2}\frac{1}{(1+q^2)^2}
  \left\{ -2qx-q^2+1\right\} \nonumber\\
  &=&  -4\,\frac{(x+q)^2 -(x^2+1)}{\Gamma_s^2\cdot(1+q^2)^2} \nonumber
\end{eqnarray}
Hence we obtain for the change $\delta\rho_s$ in the spectral density of the $s$-level
due to the coupling to the $d$-level:
\begin{equation}
  \delta\rho_s(\omega) = \frac{4\,z\,V_{sd}^2}{\pi\,\Gamma_K\,\Gamma_s^2} \cdot
  \frac{(x+q)^2 -(x^2+1)}{(x^2+1)\cdot(1+q^2)^2}
\end{equation}

Summing up, we find for the $s$-channel contribution to the transmission (\ref{eq:stransm}):
\begin{eqnarray}
  \lefteqn{
    T_s(\omega) = T_s^0 + 2\pi\cdot\frac{\Gamma_{L,s}\cdot\Gamma_{R,s}}{\Gamma_s}
  \cdot\delta\rho_s(\omega)} \nonumber\\
  &&= T_s^0 \cdot \left[ 1 + \frac{2\,z\,V_{sd}^2}{\Gamma_s\cdot\Gamma_K} \cdot
  \frac{(x+q)^2 -(x^2+1)}{(x^2+1)\cdot(1+q^2)} \right]
\end{eqnarray}
where $T_s^0=2\pi\,\rho_s^0\cdot(\Gamma_{L,s}\cdot\Gamma_{R,s})/\Gamma_s$
is the transmission via the unperturbed $s$-channel.

For the $d$-channel contribution to the total transmission we find according to eqs. 
(\ref{eq:fulltransm}) and  (\ref{eq:rhod}):
\begin{equation}
  T_d(\omega) = 4\frac{\Gamma_{L,d}\cdot\Gamma_{R,d}}{\Gamma_d^2}\cdot\frac{1}{x^2+1}
\end{equation}

Finally, for the $sd$-coupling contribution to the transmission we need to calculate 
the absolute square of the off-diagonal matrix element $G_{sd}$ of the atomic GF:
\begin{eqnarray}
  |G_{sd}(\omega)|^2 &=& V_{sd}^2\cdot |G_s^0|^2 \cdot |G_d(\omega)|^2 = \frac{V_{sd}^2\,z^2}{\Gamma_K^2\cdot(1+x^2)}
  \hspace{5ex}
\end{eqnarray}
Hence we find:
\begin{equation}
  T_{sd}(\omega) = T^0_s \cdot \frac{z^2\,V_{sd}^2}{\Gamma_K^2} \cdot 
  \left( 
  \frac{\Gamma_{R,d}}{\Gamma_{R,s}} + \frac{\Gamma_{L,d}}{\Gamma_{L,s}}
  \right) \cdot \frac{1}{1+x^2}
\end{equation}

\bibliography{Kondo_bibliography}

\end{document}